\documentclass[aps,12pt,preprintnumbers,axodraw,nofootinbib]{revtex4}
\usepackage{epsfig}
\usepackage{amsmath}
\usepackage{bm}
\usepackage{times}
\usepackage{graphicx}
\usepackage{color}
\usepackage{dsfont}
\usepackage{wrapfig}
\def\nn{\nonumber}
\def\bea{\begin{eqnarray}}
\def\eea{\end{eqnarray}}
\def\ba{\begin{eqnarray}}
\def\ea{\end{eqnarray}}
\def\be{\begin{equation}}
\def\ee{\end{equation}}
\def\beq{\begin{equation}}
\def\eeq{\end{equation}}
\def\sfrac#1#2{{\textstyle{#1\over #2}}}
\def\lsim{\mbox{\raisebox{-.6ex}{~$\stackrel{<}{\sim}$~}}}
\def\gsim{\mbox{\raisebox{-.6ex}{~$\stackrel{>}{\sim}$~}}}
\def\phm{\phantom{-}}
\newcommand{\slashed}{\slash \hspace{-0.23cm}}

\def\yht{y_t}                   
\def\yt{y_t}                    
\def\yb{y_b}                    
\def\yUD#1{{y_{#1}^{U/D}}}
\def\mH{m_{H}}                
\def\mA{m_{A}}                
\def\mHA{m_{H/A}}           
\def\SH{s_R}                    
\def\SA{s_I}                    
\def\SR{s_R}                    
\def\SI{s_I}                    
\def\SRsq{s_R^2}
\def\SIsq{s_I^2}
\def\Spm{s^\pm}
\def\DO{$\rm DO \! \! \! \!/ \, $}
\def\DOO{{\rm DO} \! \! \! \!/ \, }

\def\SHA{s_{R/I}}               
\def\GHA{\Gamma_{S_R/S_I}}      
\def\epH{\epsilon_{S_R}}        
\def\epA{\epsilon_{S_I}}        

\def\LthreeR{\lambda_3^R}
\def\LthreeI{\lambda_3^I}
\def\LfourR{\lambda_4^R}
\def\LfourI{\lambda_4^I}
\def\LfiveR{\lambda_5^R}
\def\LfiveI{\lambda_5^I}
\def\LfiveRI{\lambda_5^{R,I}}

\begin{document}
\title{Electroweak Baryogenesis in Two Higgs Doublet Models\\
	   and $B$ meson anomalies}

\author{James M.\ Cline}
\email[Electronic address:]{jcline@hep.physics.mcgill.ca} 
\affiliation{Department of Physics, McGill University, Montreal, Quebec H3A 2T8, Canada}

\author{Kimmo Kainulainen}
\email[Electronic address:]{kimmo.kainulainen@jyu.fi} 
\affiliation{Department of Physics, P.O.Box 35 (YFL),
FIN-40014 University of Jyv\"askyl\"a, Finland}
\affiliation{Helsinki Institute of Physics, P.O. Box 64,
FIN-00014 University of Helsinki, Finland}

\author{Michael Trott}
\email[Electronic address:]{mtrott@perimeterinstitute.ca} 
\affiliation{Perimeter Institute for Theoretical Physics, Waterloo, ON N2J-2W9, Canada}
\date{\today}

\begin{abstract}

Motivated by $3.9 \sigma$ evidence of a 
CP-violating phase beyond the standard model 
in the like-sign dimuon asymmetry reported by \DO, 
we examine the potential for two Higgs doublet
models (2HDMs) to achieve successful electroweak baryogenesis (EWBG) while
explaining the dimuon anomaly. Our emphasis is on the minimal flavour
violating 2HDM, but our numerical scans of model parameter space 
include type I and type II models as special cases.  We incorporate
relevant particle physics constraints, including electroweak precision
data, $b\to s\gamma$, the neutron electric dipole moment, $R_b$, and
perturbative coupling bounds to constrain the model.  Surprisingly, we
find that a large enough baryon asymmetry is only consistently achieved in a 
small subset of parameter space in 2HDMs, regardless of trying to simultaneously
account for any $B$ physics anomaly.  There is some tension between 
simultaneous explanation of the dimuon anomaly and baryogenesis, but
using a Markov chain Monte Carlo we find
several models within $1\sigma$ of the central values.  
We point out shortcomings with previous studies that reached different
conclusions.
The restricted parameter space that allows for EWBG makes this scenario highly
predictive for collider searches. We discuss the most
promising signatures to pursue at the LHC for EWBG-compatible models.

\end{abstract}

\maketitle

\tableofcontents
%
\section{Introduction}
\label{intro}
%

An experimental hint for beyond the Standard Model (SM) CP 
violation was recently reported by the \DO  \, collaboration
\cite{Abazov:2011yk,Abazov:2010hv}. The like-sign dimuon asymmetry in semileptonic $B$ decay has been
observed to deviate from the SM prediction, initially with a statistical significance of $3.2 \,\sigma$. As this paper approached completion
the statistical significance of this anomaly was reported to increase \cite{Abazov:2011yk}
to $3.9 \,\sigma$, while the central value shifted within the error band of the previous measurement.\footnote{After this paper
appeared, two new measurements were published. A new result from \DO  \cite{Abazov:2011ry}
is still consistent with the dimuon anomaly, while a new result from $\rm LHC_b$ is in tension with the  dimuon anomaly. We discuss
these results in more detail in Section \ref{collider}.}
This has motivated us to revisit electroweak baryogenesis (EWBG) in two Higgs doublet
models.\footnote{Only one linear combination of 
fields plays the role of the Higgs boson, so these should more
properly be called ``two scalar doublet models,'' but here we adhere
to the customary 
convention.} 
This measurement could indicate a new CP-violating phase
contributing to the mixing of neutral $B$ mesons or it could be a
statistical fluctuation.  Interpreting this observation as a sign of
electroweak scale new physics is supported by a pattern of deviations measured
from the SM in the $B$ sector.\footnote{The measurements of $B_s \rightarrow
J/\psi \, \phi$ and $B^- \rightarrow \tau \, \nu$ also hint at the
possibility of a new phase in $B_{s,d}$ mixing (for a recent global
fit and discussion see \cite{Lenz:2010gu}). These
deviations have the correct correlations to be part of a consistent pattern
pointing to  a new CP-violating phase. The deviation of $B^- \rightarrow \tau \, \nu$
has the most statistical significance---it is a $2.6 \, \sigma$
deviation between the SM expectation \cite{Lenz:2010gu} to the
averaged measurements of $B^- \rightarrow \tau \, \nu$ performed at
Belle and Babar \cite{Ikado:2006un,Aubert:2007xj,:2008gx,:2008ch}. See \cite{Blankenburg:2011ca} for a recent discussion of
$B^- \rightarrow \tau \, \nu$ in this model framework.}
This is also interesting as the 
measured CP violation of the SM is well known to be insufficient to generate the 
baryon asymmetry of the universe in electroweak baryogenesis (EWGB) scenarios.
In this paper, we systematically reexamine the possibility of EWGB in two 
Higgs doublet models with new sources of CP violation in light of this experimental
pattern.

In any model with multiple scalar doublets, some symmetry (usually approximate)
must be invoked to suppress flavour changing neutral currents
(FCNC's).  Frequently, treatments of multi-scalar doublet models 
impose ad-hoc discrete symmetries on the couplings of quarks and
leptons to the two doublets, following Weinberg and Glashow
\cite{Glashow:1976nt}.  In this work, we focus on the framework of
Minimal Flavour Violation  (MFV)
\cite{Chivukula:1987py,Hall:1990ac,D'Ambrosio:2002ex}. This approach allows a
new physics sector with multiple scalar doublets that is naturally
consistent with flavour constraints without such discrete symmetries.   For the purposes of testing
electroweak baryogenesis, the MFV 2HDM is sufficiently 
general that previously considered ``type I'' and ``type II'' models
can be thought of as special cases of the MFV Yukawa couplings.  Therefore our analysis 
applies to a broad class of 2HDMs, even though our emphasis is on 
the MFV framework.  Ref.\ \cite{Tulin:2011wi} has recently considered EWBG
in a 2HDM which does not fit into any of these categories, but instead
has couplings to the extra singlet that are engineered to reproduce
the \DO\ anomaly while still respecting constraints on FCNCs. 

Electroweak baryogenesis in 2HDMs has been discussed in a number of 
previous papers \cite{TZ,Funakubo,Davies,CKV,Laine:2000rm,FHS} and is the prototypical
model for EWBG, where the dynamical source of CP violation is the
simplest: it is the spatially varying phase $\theta(z)$ of the top
quark mass.\footnote{Even though the MSSM has two Higgs doublets, it
provides no such phase for $m_t$ at tree level because of the
restricted form of the couplings in the Higgs potential.}\   With the
exception of ref.\ \cite{FHS}, these papers were written before there
was a consensus on the proper way to compute the source term that
appears in the Boltzmann equations needed to solve for the baryon
asymmetry.  Thus ref.\ \cite{FHS} gives the most complete treatment to
date; yet it leaves much room for the improvements that we undertake
in the present work.  We point out a heretofore unnoticed effect that
systematically suppresses the magnitude of the  phase $\theta(z)$,
making it challenging to get a large enough  baryon asymmetry. We note
that actually solving for the bubble wall profile rather than merely
parametrizing it, as was done in \cite{FHS}, typically leads to smaller
results for the baryon asymmetry.

We also find that the particle physics constraints imposed here, but not in
previous works, significantly reduce the parameter space  that can
succeed for baryogenesis.  Using Monte Carlo methods we scan over the
full allowed parameter space of the model consistent with these
constraints, rather than restricting it in an ad-hoc manner,
as has been done in previous studies.

The present work should be regarded first as a fairly general analysis
of EWBG in 2HDMs, since as we will argue, the MFV framework is broad
enough to also encompass some popular 2HDMs (Type I and Type II) that
have been considered before.  Beyond this, we also consider whether it
is possible to simultaneously account for EWBG and the level of new CP
violation in the $B$ sector suggested by the \DO\ anomaly.  It will be
shown that the two effects are largely uncorrelated, due to the fact
that we do not find significant EWBG associated with new CP violation
coupling to $b$ quarks; rather it must exsit in the top quark Yukawa
coupling for this purpose.  In fact, we will find some tension between
simultaneously explaining the \DO\ dimuon observation and getting
sufficient baryogenesis.

The paper is organized as follows.  In section \ref{model} we describe
the 2HDM model with MFV and fitting to the dimuon anomaly.  Section
\ref{constraints} summarizes the various phenomenological and
theoretical consistency constraints that we impose when scanning the
parameter space of the model.  In section  \ref{ftep} we construct the
one-loop finite-temperature effective potential that is used to
determine properties of the electroweak phase transition.  Section
\ref{ewpt} describes the results of scanning over the model parameter
space, using a Markov chain Monte Carlo (MCMC), to find models in
which the electroweak phase transition (EWPT) is strong enough for
EWBG.   In section \ref{bau} we study the baryon asymmetry generated
by these models, finding that it tends to be somewhat below the
observed value  when the dimuon anomaly is reproduced, although
exceptions can be found.  Even when the dimuon anomaly is neglected,
only relatively rare examples exist that can produce the observed
baryon asymmetry. We then discuss in section \ref{collider} the
experimental prospects at LHC for confirming the dimuon anomaly and
the MFV two Higgs doublet model, assuming that successful EWBG occurs.
Conclusions are given in section \ref{conclusions}.

\section{Model}
\label{model}

We begin with some remarks motivating the choice of MFV as our
principle for suppressing FCNCs.  The mass scale suppressing the
operators of interest enhancing $B_{s,d}$ mixing (for perturbative couplings) 
is a few hundred $\rm GeV$ \cite{Ligeti:2010ia,Blum:2010mj}.
Models that seek to explain the $\rm DO
\! \! \! \!/ \, $ measurement with such a
mass scale are strongly constrained by flavour changing measurements
that agree well with the SM. New physics (NP) models with MFV are
naturally consistent with such flavour constraints (in the Glashow-Weinberg sense \cite{Glashow:1976nt}) as the quark
flavour group $\rm G_F =  SU(3)_{U_R} \times SU(3)_{D_R} \times
SU(3)_{Q_L}$ is only broken by the SM quark Yukawa couplings $g_{U/D}$ defined as
\begin{equation}
{\cal L}_Y=g^{~~~i}_{U~j} \, {\bar u}_{iR}  \, H^T \, \epsilon \,  Q_L^j - g^{~~~i}_{D~j} \, {\bar d}_{iR} \,H^\dagger \, Q_L^j +{\rm h.c.}. 
\end{equation} 
where
\bea
 \epsilon =\left(\begin{array}{cc} 
   0 & 1 \\
  -1 & 0
  \end{array} \right).
\eea
MFV models do allow new CP-violating phases that can lead to the $\rm
DO \! \! \! \!/ \, $ signal
\cite{Buras:2010mh,Buras:2010zm,Trott:2010iz} and can possibly supply the
required extra CP violation for EWGB.  They provide an
interesting framework for examining the relation between these
experimental anomalies and the possibility of EWGB in SM extensions.
It is also interesting to assess the feasibility of EWBG in
multi-scalar doublet models with MFV, regardless of whether the
dimuon anomaly is confirmed by future experiments.

We denote by $\rm H$ the doublet that gets a vacuum expectation value (by definition the Higgs) and by $\rm S$ the doublet that does not. One can always rotate to a 
field basis where this is the case in the models we consider. Our discussion of the model
will largely follow \cite{Trott:2010iz}. The Lagrangian in the Yukawa sector is
\bea\label{general}
{\cal L}_Y&=& g^{~~~i}_{U~j} \, {\bar u}_{iR}  \, H^T \, \epsilon \,  Q_L^j - g^{~~~i}_{D~j} \, {\bar d}_{iR} \,H^\dagger \, Q_L^j
+ Y^{~~i}_{U~j}\, {\bar u}_{iR}  \, S^T \, \epsilon \,Q_L^j - Y^{~~i}_{U~j} \, {\bar d}_{iR} \,S^\dagger \, Q_L^j+ {\rm h.c.} 
\eea
where flavour indices $i,j$ are shown and summed over and color and $\rm SU(2)_L$ indices have been suppressed.  
{\rm MFV} asserts that any {\rm NP} also has $\rm G_F$ only broken by insertions proportional to Yukawa matrices, so that 
$Y^{~~j}_{U~i}, Y^{~~j}_{D~i}$ are proportional to  $g^{~~j}_{U~i}, g^{~~j}_{D~i}$. 
One can construct allowed NP terms by treating the Yukawa matrices as
spurion fields that transform under flavour rotations as
\begin{equation}
g_U \rightarrow V_U \, g_U \, V_Q^{\dagger},~~~~~~~~~~g_D \rightarrow V_D \, g_D \, V_Q^{\dagger},
\end{equation}
where $V_U$ is an element of $\rm  SU(3)_{U_R}$, $V_D$ is an element of $\rm SU(3)_{D_R}$, and $V_Q$ is an element of $\rm SU(3)_{Q_L}$, {\it i.e.},
the Yukawa matrices  transform as $g_U \sim ({\bf 3},{\bf 1}, {\bf {\bar{3}}})$ and  $g_D \sim ({\bf 1},{\bf 3},{\bf {\bar{3}}})$ under $\rm G_F$.
MFV can be formulated up to linear order in top Yukawa insertions, or extended to a nonlinear representation of the symmetry \cite{Feldmann:2008ja,Kagan:2009bn}. 
For enhanced CP violation in $B_q$ mixing we are interested in (at least) the second order terms in the expansion of the top Yukawa in MFV. 
It is sufficient in our initial discussion to only expand to next order in insertions of $g_U$ so that
\bea\label{yuk}
Y^{~~j}_{U~i} &=& \eta_U \, g^{~~j}_{U~i} + \eta'_U \, g^{~~j}_{U~k} [(g_U^\dagger)^{k}_{~l} \, \, (g_U)^{l}_{~i}] + \cdots, \nn \\
Y^{~~j}_{D~i} &=& \eta_D \, g^{~~j}_{D~i} + \eta'_D \, g^{~~j}_{D~k} [(g_U^\dagger)^{k}_{~l} \, \, (g_U)^{l}_{~i}] + \cdots.
\eea
We decompose the second scalar doublet as  $ S^T = ({S^+}, {S^0})$,
where $S^{0} = (\SH + i \SA)/\sqrt{2}$. The scalar potential is
\bea\label{pot}
V &=& \frac{\lambda}{4} \, \left(H^{\dagger \, i} \, H_i - \frac{v^2}{2} \right)^2 + m_1^2 \, (S^{\dagger i} \, S_i)
+ (m_2^2 \, H^{\dagger \, i} S_i + {\rm h.c.}), 
\nn \\
&+& \lambda_1 \, (H^{\dagger \, i} H_i) \, (S^{\dagger \, j} S_j), 
+ \lambda_2 \, (H^{\dagger i} \, H_j) \, (S^{\dagger j} \, S_i)  + \left[\lambda_3 H^{\dagger i} \, H^{\dagger j} \, S_i \, S_j + {\rm h.c.} \right],
\nn \\
&+& \left[\lambda_4 H^{\dagger i} \, S^{\dagger j} \, S_i \, S_j + \lambda_5 S^{\dagger i} \, H^{\dagger j} \, H_i \, H_j + {\rm h.c.} \right] 
+ \lambda_6 (S^{\dagger i} S_i)^2,
\eea
where $i,j$ are $\rm SU(2)$ indices. Here $v \simeq 246 \, {\rm GeV}$ is the vacuum expectation value (VEV) of the Higgs field. Since we adopt the convention that the doublet $S$ does not get a VEV the parameters $m_2^2$ and $\lambda_5$ are related by,
\begin{equation}
\label{l5rel}
m_2^2+\lambda_5^\star{v^2 \over 2}=0.
\end{equation}
The spectrum of neutral real scalar fields consists of the Higgs scalar $h=\sqrt{2}\Re(H^0)$ another scalar field $\SH\equiv\sqrt{2}\Re (S^0) $ and a pseudoscalar $\SA\equiv\sqrt{2}\Im (S^0)$. However, these are not mass eigenstates; in the $(h, \SH,\SA)$ basis the neutral mass squared matrix ${\cal M}^2$ is
\begin{equation}
{\cal M}^2 =\begin{pmatrix}
		m_h^2  &  \LfiveR v^2 & \LfiveI \, v^2  \\
		\LfiveR v^2 & \mH^2 & 0 \\
		\LfiveI \, v^2  & 0 & \mA^2 \\
	\end{pmatrix} \,,
\end{equation}
where\footnote{We make $\lambda_3$ real by a phase rotation of $S$ with respect $H$. We also define $\lambda_4 = \LfourR + i\LfourI$ and $\lambda_5 = \LfiveR + i\LfiveI$}
\begin{equation}
m_h^2 \equiv \lambda v^2/2 \,, \qquad
\mH^2 \equiv m_S^2+\lambda_3 v^2 \qquad {\rm and} \qquad
\mA^2 \equiv m_S^2-\lambda_3 v^2
\label{eq:masses-hHA}
\end{equation}
with $m_S^2 \equiv m_1^2 + (\lambda_1 + \lambda_2) v^2/2$. Note that $m_H,m_A$ is associated with $\SH,\SA$.
The mass eigenstate field basis is denoted as $h',\SH',\SA'$ and can be expanded in terms of the original field basis as
\begin{equation}
h'   = h - \epH \, \SH - \epA \, \SA, \quad 
\SH' = \SH + \epH \, h,  \quad {\rm and}Ê\quad
\SA' = \SA + \epA \, h \,,
\end{equation}
where we defined the expansion parameters
$$
\epH \equiv \frac{v^2 \LfiveR}{\mH^2 - m_h^2}
\qquad {\rm and} \qquad
\epA \equiv \frac{v^2 \LfiveI}{\mA^2 - m_h^2} \,.
$$
The general Wilson coefficient for the relevant operator in the effective Hamiltonian  $ {\cal H}^{\rm NP}_q\simeq (V_{tq}^\star \, V_{tb})^2 \, C^{\rm NP}(m_t) \, {\bar b}^{\alpha}_R \, q^{\alpha}_L \, {\bar b}^{\beta}_R  \, q^{\beta}_L$
for perturbative $\LfiveRI \ll 1$ to leading nontrivial order in the MFV expansion is 
\begin{equation}\label{o2coefff}
C^{\rm NP}(m_t) = (\eta'_D)^2 y_t^4 \left\{ 
  \frac{(\lambda_3 + \epH\LfiveR - \epA\LfiveI)\, m_b^2}{\mH^2\mA^2} 
+ \frac{(\epH^2 + \epA^2) \, m_b^2}{2 \, v^2 \, m_h^2}
\right\}, 
\end{equation}
where $y_t \equiv \sqrt{2} m_t/v$.
In the above equation, the bottom quark mass $m_b\simeq 2.93~{\rm GeV}$ is evaluated at the top quark mass scale. Here 
the Wilson coefficient was obtained under the assumptions $ \mH, \mA  
> m_h,$  $\epH, \epA \ll 1$. In our convention the CKM factors are 
pulled out of the Wilson coefficient and appear explicitly in the
effective Hamiltonian. Note that the Wilson coefficient is
proportional to  $(\eta'_D)^2$,  not $|\eta'_D|^2$. In general this
parameter is complex, and the dependence on $(\eta'_D)^2$ introduces
an extra phase into the mixing of $B_s$ mesons through this Wilson
coefficient. 

For EWBG, we will make use of the different but analogous phase that
comes from the MFV expansion of the Yukawa couplings in the $\it top$
sector. However to explain both of these anomalies simultaneously a
number of phenomenological constraints common to the physics of both
effects must be accommodated. For example, the mass scales of the new
scalar states are relevant both for achieving successful
baryogenesis and for explaining the dimuon anomaly.

\subsection{CP violation}
Since new CP violation is a central ingredient to this work,
let us summarize the new CP-violating phases in our model.  The
new Yukawa couplings to the extra Higgs field $S$ are beyond the
standard model, and their phases cannot be removed by field
redefinitions, which have already been used to push the phases in the
SM-like Yukawa couplings (to the Higgs $H$) into the CKM matrix.
For our purposes, the relevant new couplings are to the heaviest
quarks, $t$ and $b$, the first of which is important for baryogenesis
and the second for the dimuon anomaly.

Beyond these two new phases, the Higgs potential has four complex
couplings, $m_2^2$, $\lambda_3$, $\lambda_4$ and $\lambda_5$.  Because
of our convention that $S$ does not get a VEV at zero temperature, 
eq.\ (\ref{l5rel}), $m_2^2$ and $\lambda_5$ are linearly related,
removing one of these phases.  Moreover we can do a phase rotation on
$S$ (relative to $H$) to remove an additional phase; we choose to make
$\lambda_3$ real.  Thus there are two unremovable phases in the Higgs
potential, in addition to the two new relevant phases in the $S$
Yukawa couplings.  We will show (see sect.\ \ref{sssect}) that nonzero
$\lambda_5$ suppresses the strength of the electroweak phase
transition, so its phase will not play an important role for us. The
effect of the phase of $\lambda_4$ is to allow for $S$ to be complex
inside the bubble walls of the EWPT, inducing a phase difference
$\varphi_{sh}$ between the neutral components of $S$ and $H$.  It is
the sum of $\varphi_{sh}$ and the phase of the new top quark Yukawa coupling,
denoted by $\varphi_\eta$, that will appear in the spatially-dependent
CP-violating phase in the bubble wall, $\theta(z)$;
see eq.\ (\ref{thetaeq}).

\subsection{Relations to other Two Higgs Doublet models}

We will show that only relatively rare examples exist leading to a large 
enough baryon asymmetry in this model.
We believe that paucity of models producing a large enough baryon asymmetry 
also holds for traditional type I or II two-Higgs doublet
models as we now explain. Consider a type I or II two-Higgs doublet
model where the matrix Yukawa couplings to $H_1$ and $H_2$ satisfy 
\beq
	\begin{array}{cc}  
	     g_U^2 = g_D^2 = 0, & {\rm type\ I},\\
		 g_U^2 = g_D^1 = 0, & {\rm type\ II},
	\end{array}
\label{type12couplings}
\eeq
and $g^i_{U,D}$ denotes the diagonal Yukawa couplings of the scalar fields to up or down quarks
respectively. Each of these Yukawa couplings is multiplied by $\eta_U^i$ and $\eta_D^i$. In this notation, both fields generically get VEV's in the ratio
$\langle H_2\rangle / \langle H_1\rangle = \tan\beta$.  The form
of the couplings (\ref{type12couplings}) can be insured by
$Z_2$ symmetries, but in order to get a CP-violating phase in the
Higgs sector for EWBG, these symmetries need to be broken; no unbroken symmetry distinguishes the scalar fields.

If the symmetry were exact, then its action on the
fields could be obscured by rotating to a basis in the $H_1,H_2$
space. But since the symmetry is broken, nothing inhibits us from redefining fields
to go to a new basis where $H_1$ is along the symmetry breaking
direction and $H_2$ gets no VEV.  By transforming the Yukawa couplings
accordingly, we immediately find that the broken {\rm type\ I} model
corresponds to the general MFV model with the condition $\eta_{U/D}^1
= - \eta_{U/D}^2 \, \cot \beta$. Similarly, the broken {\rm type\ II}
model corresponds to the condition $\eta_{U}^1 = - \eta_{U}^2 \, \cot
\beta$ and $\eta_{D}^2 = - \eta_{D}^1 \, \cot \beta$. We 
have set all other parameters $\eta'_U,\eta''_U, \eta'_D,\eta''_D \, +
\cdots$ for both scalar fields to zero in this discussion for
simplicity.   Our MCMC exploration of the parameter space would tend
toward  these relations between the couplings if they were favorable
toward EWBG.  The fact that they do not shows that  the flavour
structure of this model is not responsible for the difficulty
of achieving EWBG in our analysis.

This demonstrates that the MFV 2HDM is a general framework that
includes   {\rm type\ I,\,II}  models (in which the discrete symmetry
is softly broken, as is always required to get EWBG) where both scalar
fields couple to the quarks  as special cases.  For the purposes of
only achieving EWBG, in fact there is no interesting distinction
between the models, since as we will confirm the top quark Yukawa
couplings by themselves give the dominant contribution to the baryon
asymmetry.  Thus, if setting aside the question of $B$ physics, we can
set $\eta_D^i$ to zero.  Then the relations between $\eta_U^i$ to
reproduce type I or II Yukawa structure can always be satisfied for
some choice of $\tan\beta$.  To the extent that the top quark source
dominates EWBG, there is no significant difference between type I, II
or MVF models.

%

\subsection{Fitting to the $\rm DO \! \! \! \!/ \, \, $ and other $B$ sector anomalies}
When attempting to link EWBG 
to the recent dimuon anomaly, we use the fit of \cite{Lenz:2010gu} to determine the new contribution
to $B_q$-${\bar B}_q$ mixing (here $q = {s,d}$). The $\rm DO \! \! \! \!/ \, \, $  result that deviates from the SM prediction at $3.2 \sigma$ ($a_{SL}^b$) and the SM prediction ($A_{SL}^b$) are given by
\bea
a_{SL}^b &=& \frac{N_b^{++} - N_b^{--}}{N_b^{++} + N_b^{--}}, \nn \\
 &=& - (9.57 \pm 2.51 \pm1.46) \times 10^{-3}, \\
A_{SL}^b &=& (-3.10^{+0.83}_{-0.98}) \times 10^{-4}.
\eea
The number of $X \, b \, \bar{b} \rightarrow \mu^+ \, \mu^+ \, Y$ events is given by $N_b^{++}$ for example.
The quoted $a_{SL}^b$ is a combination of the asymmetry in each $B_q$, denoted  $a_{SL}^{bq}$. Each of these contributions to $a_{SL}^b$ can be expressed in terms of the 
mass and width differences ($M_{12},\Gamma_{12}$) of the $B_q$ meson eigenstates and the CP phase difference between these quantities $\phi_q$  as
\bea
a_{SL}^{bq} = \frac{|\Gamma_{12}^q|}{|M^q_{12}|} \, \sin \phi_q.
\eea 
The model we discuss involves a NP contribution that includes a new
CP-violating phase to $M^q_{12}$. 
It also allows other new CP-violating phases, all of which are interesting for our study, due to their potential to drive EWBG.
The effect of this model on $B_s$ and $B_d$ mass mixing is encoded in 
two real parameters,  $h_q>0$ and $\sigma_q$, by writing
\bea
M_{12}^q =  \left(M_{12}^q\right)^{\rm SM} + \left(M_{12}^q\right)^{\rm NP},  \quad \quad \left(M_{12}^q\right)^{\rm NP} = \left(M_{12}^q\right)^{\rm SM} \,  h_q \, e^{2 \, i \, \sigma_q}. 
\label{hqeq}
\eea 
Frequently in NP models that obey MFV one has a flavour universal phase, so that  $h_s = h_d$ and $ \sigma_s = \sigma_d$, and the best fit values are $h_q =0.255$ and $2\sigma_q=180^{o}+63.4^o$.
This scenario is argued to be a better fit to the current data then the SM in \cite{Lenz:2010gu}, which is disfavoured with a p-value
of $3.1 \sigma$.
A recent update of this measurement by  $\rm DO \! \! \! \!/ \, \, $ with $9 \, fb^{-1}$ of data finds that the significance of the deviation has increased to 
$3.9 \sigma$. We continue to utilize the global fit results when attempting to accommodate the dimuon anomaly and EWGB, 
as a new global best fit value is not available. The updated result is within the error band of the 
previous measurement that is incorporated into the global fit, so we expect our conclusions to hold for an update to the global fit incorporating the
more statistically significant deviation.

We treat perturbative QCD in the leading logarithmic approximation and evaluate the needed matrix elements of four quark operators using the vacuum insertion approximation at the
$b$ quark mass scale. 
The effective Hamiltonian is $ {\cal H}^{\rm NP}_q\simeq (V_{tq}^\star \, V_{tb})^2 \, C^{\rm NP}(m_t) \, {\bar b}^{\alpha}_R \, q^{\alpha}_L \, {\bar b}^{\beta}_R  \, q^{\beta}_L$
(see Eqn.\ (\ref{o2coefff})) where $\alpha,\beta$ are colour indices;
one finds
\begin{equation}
  \left(M_{12}^q\right)^{\rm NP}\simeq(V_{tq}^\star \, 
   V_{tb})^2C^{\rm NP}(m_t)\left(-{5 \over 24}\right) 
   \eta'  f_{B_q}^2 m_{B_q},
\end{equation}
with the Wilson coefficient defined as above. Using the results of  \cite{Trott:2010iz} we have
\begin{equation}\label{key}
 h_q \, e^{2 \, i \, \sigma_q} \simeq-{5 \over 8}\left({C^{\rm NP}(m_t) \over C^{\rm SM}(m_t) }\right) {\eta' \over \eta}.
\end{equation}
where $\eta \simeq 0.84$ is a QCD correction factor, 
and $\eta'  \simeq 1.45$. The SM contribution is
\begin{equation}
  C^{\rm SM}(m_t) =  \frac{G_F^2}{4 \, \pi^2} \, M_W^2 \,  
S\left(\frac{m_t^2}{M_W^2}\right).
\end{equation}
where $S(m_t^2/M_W^2) \simeq 2.35$ \cite{Bagger:1997gg}. 
Using these results we can scan over the allowed parameter space in
the potential parameters $\lambda_i$, couplings $\eta_i$, and the
masses of the new scalars, to search for examples consistent
with successful EWBG while fitting the dimuon anomaly.

\subsection{Higher order terms in the MFV expansion}
\label{hot}
Flavor breaking in  MFV is based on an expansion in the insertions of the spurions
$Y_U^\dagger Y_U$ and $Y_D^\dagger Y_D$.  These spurions can be
inserted between any contraction of the $Q_L$ flavour indices. Each term
in the expansion where a spurion is inserted comes with an unknown
parameter $\eta_U, \eta'_U, \eta''_U \cdots$. If these parameters
are $O(1)$, it is not appropriate to expand to only leading
order when considering the effects of  the
top Yukawa.\footnote{In the discussion above, we have expanded to
the leading order  required for the $B_{s,d}$ mixing effect that we
are interested in, avoiding these complications until this section in
for the sake of clarity.} Formally, one can treat these higher order
terms in the elegant GMFV approach of \cite{Kagan:2009bn}. In this
section, we discus higher order terms in this expansion and their
impact on the MFV 2HDM model. We adopt a less elegant,
but simpler approach than \cite{Kagan:2009bn} for our constraints.

We define new couplings,
$\mathcal{\zeta}_U,\mathcal{\zeta}'_U,\mathcal{\zeta}_D,
\mathcal{\zeta}'_D$, that resum parts of the series in the original
expansion. The couplings  $\mathcal{\zeta}_U, \mathcal{\zeta}_D$ are
defined as the sum of all terms in the usual MFV expansion that do not
involve flavour change, while the couplings $\mathcal{\zeta}'_U,
\mathcal{\zeta}'_D$ are defined as the sum of all terms in the
expansion in $\eta_i \, (\sqrt{2} m_t/v)^2$  that do lead to a
particular flavour change. For couplings to neutral scalar fields this  $t
\rightarrow j$ flavour change is from one charge $+2/3$ quark species to another.
For couplings to the charged scalar fields, the flavour change is from an up type
quark to a down type quark. In each case, in MFV, the transition is
accompanied by the appropriate CKM suppression of the standard model.
Then for the neutral scalar fields one has for the top couplings
\bea
\mathcal{\zeta}^0_t 
&=& \yt \left( \eta_U + \eta_U' \, \yt^2 
             + \eta_U'' \, \yt^4 + \cdots \right), \nn \\
\mathcal{\zeta}^{'0}_t &=& (V_{bt}^\star \ V_{bj}) \, 
    \yt \left(\tilde{\eta}'_U \, \yb^2 
             + \tilde{\eta}''_U \, \yb^4 + \cdots \right),
\eea
where $y_{t,b} = \sqrt{2}m_{t,b}/v$.
For flavour change one must insert $g_D^\dagger \, g_D$ spurions as a correction to the $g_U$ coupling as flavour changing effects require the
presence of both the up and down Yukawa matrices.  These insertions,
since they arise from a different spurion insertion, come with different parameters which we denote $\tilde{\eta}'_U,\tilde{\eta}''_U$ in the expansion of  $\mathcal{\zeta}^{'0}_t$.
There are also $g_D^\dagger \, g_D$ insertions in the expansion of $\mathcal{\zeta}^0_t$, but we neglect them in this expression.
For the charged fields one has for the top couplings generating the $t \rightarrow b$ flavour change 
\bea
\mathcal{\zeta}^\pm_t    &=&0, \nn \\
\mathcal{\zeta}^{'\pm}_t &=& (V_{bt}) \; \yt \left(\eta_U + \eta_U' \, \yt^2 + \eta_U'' \, \yt^4 + \cdots \right).
\eea
There are also $g_D^\dagger \, g_D$ insertions in the expansion of $\mathcal{\zeta}^{'\pm}_t$ exactly as in the expansion of $\mathcal{\zeta}^0_t$.
For the purposes of this discussion we neglect terms subleading in CKM
 or small Yukawa contributions to the effective couplings. 
The effective charged coupling multiplying the CKM insertion is the same as the effective coupling for the neutral field when no flavour change occurs.
Furthermore $V_{tb} \sim 1$ to a good approximation.
This identification allows one to directly relate the requirements of EWGB to the particle physics constraints that we impose. 
The procedure above can be repeated for the bottom quark couplings by
defining $\mathcal{\zeta}^0_b,\mathcal{\zeta}^{'0}_b,
\mathcal{\zeta}^\pm_b, \mathcal{\zeta}^{'\pm}_b$. 

For the dimuon anomaly, we are interested in the effective flavour
changing couplings involving the bottom quark for the new neutral
scalar fields. We approximated this coupling with only
the leading term in the previous section as $(V_{tb}^\star \ V_{ts})
\, \eta_D' \, y_b \, y_t^2$. It is explicitly given by
\bea
\mathcal{\zeta}^{'0}_b &=& (V_{tb}^\star \ V_{ts}) 
\yb \left( \eta'_D \, \yt^2 + \eta''_D \, \yt^4 + \cdots \right).
\eea
so that the full result for the Wilson coefficient of interest for enhanced $B_s$ mixing is given by 
\bea
C^{\rm NP}(m_t) = \left[\frac{\mathcal{\zeta}^{'0}_b}{V_{tb}^\star \ V_{ts} \, y_b} \right]^2 \, \tilde{C}.
\eea
where $\tilde C$ is the quantity in curly brackets in Eqn.~(\ref{o2coefff}).
The expression $\mathcal{\zeta}^{'\pm}_b$ appears in the
precision measurement constraint formulae in the next section.
When considering this scenario and fitting to the dimuon anomaly we 
will take 
\beq 
 \mathcal{\zeta}^{'\pm}_b/(V_{bt}) = f\,
\mathcal{\zeta}^{'0}_b/(V_{tb}^\star \ V_{ts}) 
\label{generic}
\eeq
with $f=O(1)$ as a generic relation
between $\mathcal{\zeta}^{'0}_b$, which controls the strength of the
dimuon anomaly, and $\mathcal{\zeta}^{'\pm}_b$ which appears in the
neutron EDM and $b\to s\gamma$ constraints, as explained in section
\ref{constraints}.  The motivation for 
this choice is that these parameters get the same contributions from
all the primed terms in the MFV expansion, while $\eta_D$ only 
appears in $\mathcal{\zeta}^{'\pm}_b$.  In the absence of fine-tuning
(or some principle provided by the unknown UV completion of our
theory), we expect this relation to hold with $f\sim 1$.  However if $\eta_D$ happens
to partially cancel the other contributions to 
$\mathcal{\zeta}^{'\pm}_b$, or if the addition of all the spurion 
insertion terms sums to a suppressed value (compared
to $\eta_D$), it is possible
to get a relaxation of the loop-generated constraints relative to 
the $B$-mixing contribution, hence smaller values of
$f$.  We will explore the dependence on $f$ in the subsequent analysis.

Aside from taking $f\ll 1$, another way of increasing the $B$-mixing contribution without 
simultaneously increasing the unwanted $b\to s\gamma$ and EDM
contributions is to increase $|\lambda_5|$, which causes mixing between
the scalar CP eigenstates.  However this is not helpful in the current
situation.   The problem 
is that the part of the function $\tilde{C}$ that can be large enough to generate  the dimuon
anomaly (with smaller
$\eta_D'$) is
proportional to $(\LfiveRI)^2$ and the first order phase transition
required for EWBG  strongly prefers $|\LfiveRI| \ll 1$. In physical
terms, the mass mixing between $H$ and $S$ that is driven by
$\lambda_5$ to allow the dimuon anomaly to be fit to (with smaller
$\eta_D'$) directly suppresses the first order phase transition
required for EWBG. This is discussed in more detail in Section
\ref{bau}. If $|\LfiveRI| \ll 1$, the mass and VEV eigenstates
of the scalar fields approximately coincide, which makes discovery of
the new scalar states challenging above the $t \, \bar{t}$ threshold as we will discuss in Section 
\ref{collider}.

%
\section{Phenomenological Constraints}
\label{constraints}
%

In this section we discuss  the constraints on the 2HDM model that we impose when searching for parameter space
with viable EWBG.
The constraints are enumerated in Table I, which specifies the
model parameters that are most directly affected by each one for the reader who wishes to skip the details.
\begin{table}[h]
\begin{center}
\begin{tabular}[t]{|c|c|c|c|c|c|}
\hline
\hline
Constraint    
        &  Constrained  Parameters \\
  \hline
{\rm LEP/Tevatron Direct Search}   
        & $m_1, \lambda_1,\lambda_2,  \mathcal{\zeta}^{'\pm}_t$ \\
{\rm RGE/Landau Pole/Unitarity} \, 
        & $\lambda, m_1, \lambda_i,  \mathcal{\zeta}^{0}_t$ \\  
{\rm Neutron EDM} \, 
        & ${\rm Im} [(\mathcal{\zeta}^{'\pm}_t)^\star \,  
          (\mathcal{\zeta}^{'\pm}_b)^\star ], m_\pm(m_1, \lambda_{1,2,3})$ \\ 
$b \rightarrow s \, \gamma$ \, 
        & ${\rm Re} [(\mathcal{\zeta}^{'\pm}_t)^\star \,   
          (\mathcal{\zeta}^{'\pm}_b)^\star ], |\mathcal{\zeta}^{'\pm}_t|^2, 
          m_\pm(m_1, \lambda_{1,2,3})$ \\ 
$R_b$ \, 
        & $|\mathcal{\zeta}^{'\pm}_t|^2 \, |\mathcal{\zeta}^{'\pm}_b|^2,  
          m_\pm(m_1, \lambda_{1,2,3})$ \\ 
{\rm EWPD} \, 
        & $m_1, \lambda_{1,2,3}$\\
\hline
\hline
\hline
\end{tabular}
\end{center}
\caption{ Summary of constraints.}
\label{int-scalar}
\end{table}

%
\subsection{Collider Mass bounds}
%

The kinematic direct production bound from LEP
demands that the sum of new scalar and pseudoscalar Higgs masses obeys
$\mH + \mA > 209$ GeV. 
This is consistent with the assumption $\mH, \mA > m_h$ made
in deriving the Wilson coefficient in Eqn. (\ref{o2coefff}).
Moreover
\bea
\sfrac12(\mH^2+\mA^2) = m_1^2 + \frac{\lambda_1 + \lambda_2}{2} \, v^2 > m_h^2.
\eea
The remaining direct production bound from LEP is on the 
charged scalar mass, $m_\pm = \sqrt{m_1^2 + \lambda_1/2} > 105$ GeV. 
This bound is a purely kinematic constraint not dependent 
on tagging particular final states.\footnote{The quoted bound of $78.6  \, {\rm GeV}$ at $95 \% {\rm CL}$ in \cite{:2001xy} is specific to the case ${\rm Br} \left(S^+ \rightarrow \bar{s} \, c\right) + {\rm Br} \left(S^+ \rightarrow\tau^+ \,  \nu_\tau \right) = 1$. We impose the 
more conservative kinematic bound.}

For masses $m_\pm < m_t$ the constraints from the Tevatron can also be
used. The CDF collaboration \cite{Aaltonen:2009ke} constrains ${\rm
Br} \left(t \rightarrow S^+ \, b \right)$ through subsequent decays of 
$S^+ \rightarrow c \, \bar{s}$, while the $\rm
DO \! \! \! \!/ \, $ collaboration \cite{Abazov:2009wy} uses the
subsequent decays $S^+ \rightarrow \tau^+ \,
\nu_\tau$. The latter decay involves the coupling of the
charged scalar to leptons, which is  not directly related to the
parameters of interest. Therefore we use the CDF  result which for
$m_s^\pm < 150 \, {\rm GeV}$ gives the constraint
\bea
{\rm Br} \left(t \rightarrow S^+ \, b \right){\rm Br} \left(S^+ \rightarrow c \, \bar{s} \right) \lesssim 0.1.
\eea
For the 2HDM model with $m_\pm < 150 \, {\rm GeV}$ we have ${\rm Br} \left(S^+ \rightarrow c \, \bar{s} \right) \sim 1$ and this becomes
\bea
\frac{|\mathcal{\zeta}^{'\pm}_t|^2 |V_{tb}|^2}{8 \, \pi} \, \frac{(m_t^2 - (m_s^\pm)^2)^2}{m_t^3 \, \Gamma_t} \lesssim 0.1
\eea
where $\Gamma_t \sim 1.3 \ {\rm GeV}$.

\subsection{Consistency Conditions on the Potential}

Although renormalizable, for our purposes this minimal model is best
thought of as an effective low energy scalar sector with field content
of a UV completion at a relatively low scale $\Lambda \sim \, {\rm
TeV}$. We will find that $\mathcal{O}(1)$ parameters are required in
the potential which can lead to a relatively low cutoff scale. For
consistency we ensure that the couplings of the allowed parameter
space do not approach a Landau pole or cause the potential to  be
unbounded from below up to $\Lambda \sim 1$-$2 \, {\rm TeV} \gg m_i$,
where $m_i$ are the masses in the two scalar doublet model. We impose
this constraint considering running and vacuum stability under the
complete one-loop RGE flow of the couplings in the potential.

The RGE running of the potential parameters can be derived directly
by modifying the effective potential method of 
\cite{Ferreira:2009jb}. For complex coefficients,  within the
convention that $S$ is rotated so that $\lambda_3$ is real, the scaled
beta functions are given in Appendix \ref{app:betafns}. 
Here the one-loop beta
functions are defined in terms of the functions defined above as
$\beta_x = \hat\beta_{x}/16 \, \pi^2$. In these expressions, $g'$ 
is the
$\rm U(1)$ gauge coupling $g$ is the $\rm SU(2)$ coupling
of the standard model. The top Yukawa 
coupling of the Higgs field, and
the complex Yukawa couplings of the new doublets are defined as
\bea
\yht = \sqrt{2} \, \frac{m_t}{v}, \quad \quad \eta_t = \mathcal{\zeta}^{0}_t, \quad \quad \eta_b =  \mathcal{\zeta}^{0}_b.
\eea
We also need the standard results of the one-loop running of the gauge couplings (above $m_t$) and the Yukawa couplings; they are
\bea
\hat\beta_{g'} = 7 g'^3, \quad \quad \hat\beta_g &=& - 3 g^3,  \quad \quad
\hat\beta_3 = -7 g_3^3, \quad \quad \hat\beta_{\yht} =  \yht \left[\frac92 \, \yht^2- \frac{17}{12} g'^2 - \frac94 g^2 - 8 g_3^2 \right], \nn \\
\hat\beta_{\eta_{t,b}} &=&  \eta_{t,b} \left[\frac92 \, |\eta_t|^2 + \frac92 \, |\eta_b|^2- \frac{17}{12} g'^2 - \frac94 g^2 - 8 g_3^2 \right].
\eea
Here we have neglected small mixing effects of the Yukawa coupling operators which we discuss below.
For vacuum stability we assume that $\lambda_6 > 0$,
which is sufficient for the field space direction along the $S$ axis, considering the masses are constrained to be positive and large due to the bounds discussed above.
For directions not along the $h$ or (a particular) $S_i$ axis, it is
difficult to analytically formulate the conditions for stability
when $\lambda_{4,5}\neq 0$, but it is easy to numerically check
for runaway behavior, and this is the approach we take.

{}From \cite{Ginzburg:2005dt} we also have the unitary scattering
constraint. As we scan parameter space and evolve the couplings of the
model under the RGE we insist that the theory remain unitary as the
couplings are evolved for mass scales up to the cutoff scale of the
theory. However, we find the unitarity constraint has a negligible
impact on the parameter space of interest; it is weaker than the
demand to avoid Landau poles.

\subsection{Phenomenological constraints on the loop corrections 
due to charged scalars}

 In this section we consider the constraints arising from
virtual scalar exchange contributions to the neutron EDM, $Z\to\bar b
b$ and $b\to s\gamma$.  In all of these processes, the dominant
contribution comes from charged $S^\pm$ exchange because of the fact
that the flavor change $t\to b$ in the loop is not CKM suppressed.

\subsubsection{Neutron EDM}
The effect of the exchange of virtual charged and neutral scalars on
precision observables leads  to significant constraints on models
that would otherwise have given rise to successful EWBG.  
Some electric dipole moment (EDM) constraints on this
scenario were discussed in \cite{Buras:2010zm}. We focus on the EDM
constraint that is not suppressed by small mixing angles or light
quark masses that was discussed in \cite{Trott:2010iz}. For the
neutron EDM we have
\bea
\left|{\rm Im }\left[\frac{v^2 \,  (\mathcal{\zeta}^{'\pm}_t)^\star \,
(\mathcal{\zeta}^{'\pm}_b)^\star }{2 \, m_t \, m_b}\right]\right|\, 
f_g \left(\frac{m_t^2}{m_\pm^2}\right)
< 0.043, \quad  f_g(x) = \frac{x\log(x)}{(x -1)^3}  + \frac{x(x - 3)}{2 \, (x -1)^2}.
\eea
Naive Dimensional Analysis (NDA) is used to compute the neutron EDM
matrix element in this estimate.\footnote{Although the uncertainty in
the hadronic matrix element is significant and alternate estimates 
such as in \cite{Demir:2002gg} can weaken this constraint,  we
conservatively use the NDA estimate in restricting the parameter
space.}  To satisfy this constraint one generically has two choices:
 the parameter
$\mathcal{\zeta}^{'\pm}_t$ can be small, or the relative phases of
$(\mathcal{\zeta}^{'\pm}_t)^\star$ and 
$(\mathcal{\zeta}^{'\pm}_b)^\star$ can be tuned so that the
constraint is satisfied.\footnote{In general
$(\mathcal{\zeta}^{'\pm}_t)^\star$ and
$(\mathcal{\zeta}^{'\pm}_b)^\star$ are the combination of many terms
that can each individually have an independent phase in the MFV
expansion.} Either of these choices could conceivably be justified by further model
building in the UV.

\begin{figure}[tp]
\centerline{\scalebox{0.6}{\epsfig{file=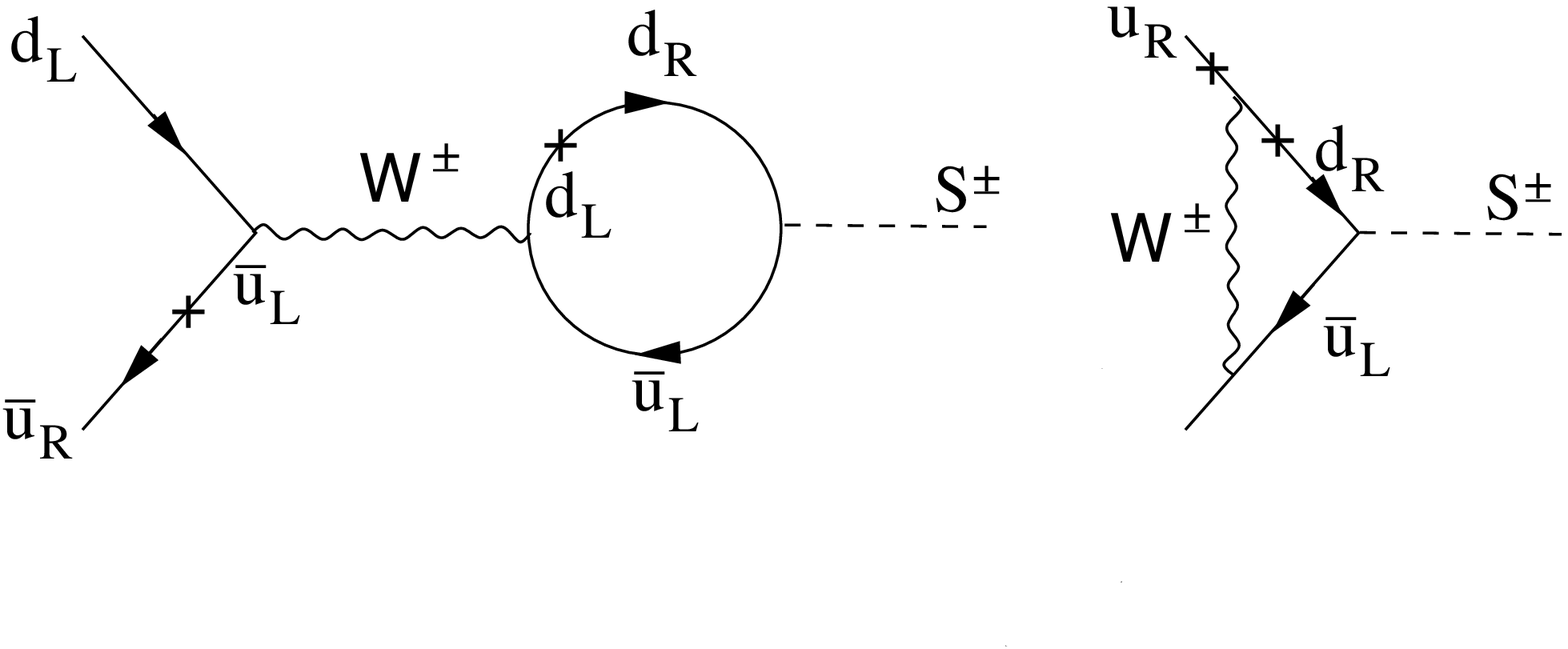}}}
\vskip-2cm
\caption{One loop diagrams that induce an effective local operator of the charged scalar field $\Spm$ to $\bar{u}_R \, d_L$ from a coupling to $\bar{u}_L \, d_R$. The largest
contribution for the diagrams come from a top-bottom quark fermion 
loop in both cases.}
\label{fig1}
\end{figure}

Taking the parameter $\mathcal{\zeta}^{'\pm}_t$ small to satisfy the
EDM bound generally requires $|\mathcal{\zeta}^{'\pm}_t|  \lesssim
10^{-1}$ and the allowed $\mathcal{\zeta}^{'\pm}_t$ decreases as
$\mathcal{\zeta}^{'\pm}_b$ increases. One can also accomplish the
suppression of $\mathcal{\zeta}^{'\pm}_t$ naturally by using MFV in a
model where the new scalar doublet is not a flavour singlet; see
\cite{Trott:2010iz}. However, there is no exact symmetry in models of
this form that distinguishes between the coupling of the charged
scalars to $\bar{u}_R \, d_L$ and $\bar{u}_L \, d_R$;  thus the
effective parameters $\mathcal{\zeta}^{'\pm}_t$ and
$\mathcal{\zeta}^{'\pm}_b$ are not independent and the corresponding
operators mix. Therefore the radiative
stability of such a choice is an interesting issue. An
effective $\mathcal{\zeta}^{'\pm}_t$ is induced through the one-loop
diagrams in Fig.\ \ref{fig1} from the coupling of the charged scalars to
$\bar{u}_L \, d_R$  proportional to
$(\mathcal{\zeta}^{'\pm}_b)^\star$.
%
%
%
%
Calculating these diagrams one finds a contribution to the  effective $\bar{u}_R \, d_L$ Yukawa coupling such that
\bea
\delta \mathcal{\zeta}^{'\pm}_t &\approx& \frac{ (\mathcal{\zeta}_b^{'\pm})^\star \, y_b}{8 \, \pi^2} \, \int_0^1 \,dx \, (1-x) \, \log \left[\frac{\mu^2}{x \, m_t^2 + (1-x) m_b^2} \right], \nn \\
&\,&  + \frac{(\mathcal{\zeta}_b^{'\pm})^\star  \, y_t}{128 \, \pi^2} \, \int_0^1 dx \, \int_0^{1-x} dz \left(-1 + (1 + x +z) \, \log \left(\frac{\mu^2}{\Delta} \right) \right), \nn \\
&\,&+ \frac{(\mathcal{\zeta}^{'\pm}_b)^\star  \, y_t }{8 \, \pi^2} \, \int_0^1 dx \, \int_0^{1-x} dz \, (1 - x -z) \,  \left[\frac{m_t^2 \, \left(x+ z \right)^2 }{\Delta} +  \frac{1}{2} \left(1 - \log \left(\frac{\mu^2}{\Delta} \right) \right)\right].
\eea
Here $\Delta = - m_t^2 ( x (1-x) - z^2 - 4 x z - z) + m_b^2 (2 x z + 2 z^2 - z + x) - (1-x -z)M_W^2$ and we have retained the finite terms of these diagrams to illustrate the effect. For $\mathcal{\zeta}^{'\pm}_b \gtrsim m_t/m_b$ 
this contribution to the effective Yukawa coupling is large enough
so that the induced effect on $b \rightarrow s \, \gamma$ and the neutron $\rm EDM$ can be in conflict with experiment even if the coefficient of the operator  $\bar{u}_R \, d_L$ is chosen to be small at tree level.
For this reason we restrict our choice of coupling to $|\mathcal{\zeta}^{'\pm}_b| \lesssim 10$ when scanning parameter space.  

For EWBG, it is advantageous to satisfy the EDM bound by tuning the
relative phases.  In this case  $\mathcal{\zeta}^{'\pm}_t  >
\mathcal{\zeta}^{'\pm}_b$ and one can have a strong coupling of the
new Higgs $S$ to the top quark.   If on the other hand we were to
suppress $\mathcal{\zeta}^{'\pm}_t  \sim 10^{-2}$ using the MFV model
of \cite{Trott:2010iz} or by choosing $\mathcal{\zeta}^{'\pm}_t$
to be small,  the small effective top quark coupling to the new field
(and new CP violation) suppresses EWBG.

\subsubsection{$Z \rightarrow b \, \bar{b}$ constraints}

Another important  constraint comes from limits on the charged scalar
masses and couplings due to modifications of the $Z\, b \, \bar{b}$
vertex. $Z \rightarrow b \, \bar{b}$ provides direct bounds on the
neutral and charged scalars.\footnote{ See
also \cite{Degrassi:2010ne,Jung:2010ik} for recent discussions on
$R_b$ constraints in models of this form focused on charged scalar
exchange.}\ \  The shift in $R_b$ due to the virtual corrections of the
new scalars comes about through the shift in $\bar{g}_b^R =
(\bar{g}_b^R)_{SM} +  \delta \bar{g}_b^R$, $\bar{g}_b^L =
(\bar{g}_b^L)_{SM} +  \delta \bar{g}_b^L$ for the $b$ quark to the $Z$
and we parameterize this effect on $R_b$ by 
\bea
\delta R_b \simeq 2 \, R_b \, ( 1- R_b) \, \left(\frac{\delta \bar{g}_b^L \, (\bar{g}_b^L)_{SM} + \delta \bar{g}_b^R \, (\bar{g}_b^R)_{SM}}{(\bar{g}_b^L)_{SM}^2 + (\bar{g}_b^R)_{SM}^2} \right)
\eea
We use the predicted $\rm SM$ values $(\bar{g}_b^R)_{SM} = 0.0774$, $(\bar{g}_b^L)_{SM} = -0.4208$ which gives $R_b = 0.21578 \pm 0.00010$.
Using these results we have 
\bea
\delta R_b \simeq - 0.78 \, (\delta \, \bar{g}_b^L)  + 0.14 \, (\delta \, \bar{g}_b^R).
\eea
Considering that the measured value given by the Particle Data Group is $R_b = 0.21629 \pm 0.00066$  
\cite{PDG} we have the $1.3 \, \sigma$ bounds
$ - 1.6 \, \times 10^{-4} < \delta \, R_b <  1.1 \, \times 10^{-2}$.
The shifts in the couplings, where we neglect suppressed corrections
of the form $\eta_i \, \lambda_j/16 \, \pi^2$ due to mixing and
$m_b^2/\mA^2,M_Z^2/\mA^2$ suppressed terms are given by
\bea
\delta \bar{g}_b^L &\simeq &  \frac{|\mathcal{\zeta}^{'\pm}_t|^2}{32 \, \pi^2}  \left(\frac{m_t^2/m_\pm^2}{m_t^2/m_\pm^2 - 1}- \frac{m_t^2/m_\pm^2  \log \left(m_t^2/m_\pm^2 \right)}{(m_t^2/m_\pm^2 - 1)^2} \right),  
\nonumber \\
\delta \bar{g}_b^R &\simeq& 
- \frac{|\mathcal{\zeta}^{'\pm}_b|^2}{|\mathcal{\zeta}^{'\pm}_t|^2}\,\delta\bar{g}_b^L \,.
\eea
Here $s_W,c_W$ are the sin and cosine of the weak mixing angle. We choose the renormalization scale $\mu = M_Z$.

\subsubsection{Constraints on $\mathcal{\zeta}^{'\pm}_t$ through $b \rightarrow s \, \gamma$}

As explained above, we are interested in $\mathcal{\zeta}^{'\pm}_t 
\sim 1$ to allow for EWGB, and so we must include terms in the $b \rightarrow s \, \gamma$ constraint
proportional to $(\mathcal{\zeta}^{'\pm}_t)^2$. The constraint from $BR(\bar{B} \rightarrow s \, \gamma)_{E_\gamma> 1.6 {\rm GeV}}$ at $95 \%$ CL \cite{Grzadkowski:2008mf} is given by
\bea
-0.42 < -{\rm Re} \left[(\mathcal{\zeta}^{'\pm}_t)^\star \, (\mathcal{\zeta}^{'\pm}_b)^\star \right]
f^{(2)}_\gamma\left(\frac{m_t^2}{m_\pm^2}\right)   
+ \sfrac13 |(\mathcal{\zeta}^{'\pm}_t)|^2\,
 f^{(1)}_\gamma\left(\frac{m_t^2}{m_\pm^2}\right)   < 0.12,
\eea
where from \cite{Ciuchini:1997xe} we have
\bea
 f^{(1)}_\gamma(x) &=& \frac{x(7 - 5 x - 8 x^2)}{8(x-1)^3} + \frac{3 x^2 (3 x
- 2)}{4(x-1)^4} \, {\rm log}(x), \\
 f^{(2)}_\gamma(x) &=& \frac{x(3 - 5 x)}{4(x-1)^2} + \frac{ x(3 x -
2)}{2(x-1)^3} \, {\rm log}(x).\nn
\eea

\subsection{Electroweak precision data}

As we scan parameter space we also restrict the mass spectrum of the
scalars to be within the $95 \% \, {\rm CL}$ region of an electroweak
precision data (EWPD) fit. 
The oblique parameters we use, due to the low mass scale involved, are
the STUVWX parameters of \cite{Burgess:1993mg,Maksymyk:1993zm}. The
one loop corrections to the EWPD observables for the 2HDM are standard and are given in \cite{Trott:2010iz} (with
the appropriate rescaling).  The fit we use is described in
\cite{Burgess:2009wm}. We float the Higgs mass at one loop while
testing the parameter set against the EWPD constraints and perform a
joint fit to the Higgs mass and the new scalar spectrum. To do this we
use the self-energy corrections due to the Higgs at one loop given by
\bea
\Pi_{WW}(p^2) &=& \frac{g_1^2}{16 \pi^2} 
\left[- \frac{A_0(m_h^2)}{4}  
      - M_W^2 \, B_0(p^2,m_h^2,M_W^2) 
     + B_{22}(p^2,m_h^2,M_W^2) \right], 
\nn \\
\Pi_{ZZ}(p^2) &=& \frac{g_1^2}{16 \pi^2\cos^2 \theta_W} \left[- \frac{A_0(m_h^2)}{4}  - M_Z^2 \, B_0(p^2,m_h^2,M_Z^2) + B_{22}(p^2,m_h^2,M_Z^2)\right].
\eea
The one-loop functions used here are defined in \cite{Burgess:2009wm} as are the STUVWX parameters.

This test is most sensitive to the mass splitting in the scalar
spectrum.  As the splitting violates custodial symmetry, it leads
to important constraints on $\lambda_3$.  For the MCMC scans
we numerically interpolate from a grid scan of EWPD using the 
$\chi^2$ measure defined through the cumulative distribution function
for a six-parameter fit.  When we float the Higgs mass and scan the
parameter space we interpolate between the discrete masses $m_h = \{115,130,145,160\} \,
{\rm GeV}$. For the new scalar masses  we require $\mH, \mA, m_\pm
\leq 700 \, {\rm GeV}$ and vary the masses in this range over the
allowed region  considering all other consistency constraints on the
scalar potential.  The MCMC does not show a preference for masses 
exceeding this 700 GeV upper bound.

Larger Higgs masses $m_h$ are allowed in principle in these joint fits 
due to the positive $\Delta T$ contribution from the mass splitting 
in the scalar spectrum. We find that successful EWBG prefers a light Higgs.

\section{Finite temperature effective potential}
\label{ftep}
The properties of the electroweak phase transition and the bubble wall profiles needed for computing the baryon asymmetry require knowledge of the effective potential $V_{\rm eff}(H,S)$ at finite temperature. We compute it at one-loop order, along with the corresponding zero-temperature loop correction.  The full potential can be expressed as
\beq
V_{\rm eff} = V_{\rm Tree} + V_{\rm CW} + V_{\rm CT} + V_{\rm T} 
\eeq
where $V_{\rm Tree}$ is given by (\ref{pot}), $V_{\rm T}$ is the thermal
contribution, $V_{\rm CW}$ is the one-loop Coleman-Weinberg (CW) potential,
and $V_{\rm CT}$ are counterterms, which for convenience can be chosen so
as to maintain various tree-level relations for the scalar mass
eigenvalues and VEVs.  

The CW potential is given by
\beq
	V_{\rm CW} = \sum_i \pm {g_i\over {64\pi^2}} m^4_i\left(
	\ln {m^2_i\over Q^2} - \frac32\right)
\eeq
where $g_i$ is the multiplicity of species $i$, $m_i$ is its field
dependent mass and + sign is for bosons and $-$ sign for fermions. 
The field-dependent masses are given in Appendix \ref{masses}. $Q$ is a renormalization scale, which we take to be $v/\sqrt{2}$, where $v=246$ GeV is the Higgs VEV.
Some residual $Q$-dependence remains in our results since we do not perform a complete renormalization at 1-loop level including wave function renormalization. We include the Higgs bosons and electroweak gauge bosons in the sum, as well as the top quark.  We work in Landau gauge where the Fadeev-Popov ghosts decouple at the one-loop level.  

The effective potential is a function of two complex fields, $H^0$ and $S^0$, but we have the freedom to remove the phase of $H^0$ (the Goldstone boson mode) by making the appropriate SU(2) gauge transformation on both fields.  Doing so simplifies our task by eliminating the unphysical degree of freedom.  We thus regard $V_{\rm eff}$ as a function of the three fields $h,\SR,\SI$ as defined below Eqn.~(\ref{l5rel}): $H^0 \equiv \sfrac{1}{\sqrt{2}} h$ and $S^0  \equiv \sfrac{1}{\sqrt{2}}(\SR + i \SI)$.

\subsection{Counterterms and Goldstone bosons}
For convenience we introduce counterterms to preserve the tree-level relationships for masses and VEVs in the zero-temperature potential.  In the \{$h, \SR ,\SI$\} field basis we thus demand that
$\partial V/\partial h = \partial V/\partial \SR =  \partial V/\partial \SI = 0$ at $h = v$, $\SR=\SI=0$.  Denoting the tree-level VEVs by $\langle \phi_i\rangle$, this requires
\beq
\left.{\partial V_{\rm CT}\over \partial
\phi_i}\right|_{\langle \phi_i\rangle} = 
-\left.{\partial V_{\rm CW}\over \partial
\phi_i}\right|_{\langle \phi_i\rangle}
\label{vct}
\eeq
Similarly, to preserve the mass relations we require that
\beq
\left.{\partial^2 V_{\rm CT}\over \partial\phi_i\partial \phi_j}\right|_{\langle \phi_i\rangle} = 
-\left.{\partial^2 V_{\rm CW}\over \partial\phi_i\partial \phi_j}\right|_{\langle \phi_i\rangle} \,,
\label{mct}
\eeq
where $\phi_i = h, \SR, \SI$. There is a problem in principle
with carrying out (\ref{mct}). We work in the Landau gauge where
ghosts decouple but Goldstone boson (GB) contributions are retained in
the sum over species. In particular, intermediate GBs contribute to
the mass of the Higgs, and formally their contribution to $\partial^2
V_{\rm CW}/\partial h^2$ is IR log divergent due to terms of the form
$(\partial m^2_{gb}/\partial h)^2 \ln m^2_{gb}$, where the prefactor
does not vanish at the VEVs, but the argument of the log does.
This shows that renormalizing the Higgs mass at zero external
momentum with massless Goldstone modes is not a well defined
procedure. Some authors choose to simply omit the GB contributions to
the effective potential, but in the present study we find that this is
not justified.  Of course one can renormalize
the Higgs mass at any momentum scale $p^2 \neq 0$ and a consistent
procedure of renormalizing on-shell at $p^2=m_h^2$ was implemented in
ref.\ \cite{Cline-Lemieux}. This cures the GB-problem because the external Higgs
on-shell momentum flows through the loop of internal GBs contributing
to the Higgs self-energy and cuts off the IR divergence that occurs 
when renormalizing at $p^2=0$, Eqn.\  (\ref{mct}). The effect can be
approximately captured by replacing $m^2_{gb}\to m^2_{\rm IR}$ in
(\ref{mct}) for the determination of the counterterms, where $m^2_{\rm
IR}$ is some IR cutoff. Since the IR divergence is only logarithmic,
the effective potential is not greatly sensitive to the exact value. 
We find that $m^2_{\rm IR}\cong m_h^2$  gives a good approximation to
the more exact prescription of ref.\ \cite{Cline-Lemieux}.  This is
the procedure we adopt. 

In practice, not all possible counterterms are required. This is
because we do not expect to be able to measure all of the couplings in
the scalar potential in the near future, and therefore
we do not need to relate all of them (such as $\lambda_6$) to
observables. For our purposes, it is sufficient to include nine
counterterms,
\bea
	V_{\rm CT} 
	&=& \sfrac{1}{16}\delta\lambda\, h^4 - \sfrac12 \delta\mu^2\, h^2 
	    +\sfrac12 \delta m_1^2\, (\SRsq+\SIsq) 
	    + h\,(\delta m^2_{2R}\,\SR - \delta m^2_{2I}\,\SI)
\nonumber\\
	&+& \sfrac12 \delta \LthreeR h^2(\SRsq-\SIsq)
	- \delta \LthreeI \, h^2\, \SR\, \SI + \sfrac12
h^3(\delta\LfiveR\,\SR + \delta\LfiveI\,\SI) \,,
\eea
which are fixed by the nine independent renormalization conditions set
by eqs.~(\ref{vct}-\ref{mct}). Explicit formulas for the counterterm
couplings defined in this way are given in Appendix \ref{ctapp}.

%
\subsection{Thermal corrections} 
%
%
The unimproved pure thermal contribution to $V_{\rm eff}$ is given by: 
\beq
	V_{\rm T} = {T^4\over 2\pi}\sum_i \pm g_i \int_0^\infty dx\, x^2\,
	\ln\left(1 \mp \exp(-\sqrt{x^2 + m_i^2(H^0,S^0)/T^2}\right) \,,
\label{VT}
\eeq
where upper signs are for bosons and lower signs for fermions. Splitting Eqn.~(\ref{VT}) into distinct sums over bosonic and fermionic species and 
expanding it to order $O(m_i^6/T^2)$ one finds (see for example ref.\
\cite{Arnold-Espinosa}): 
\bea
V_T \cong c T^4 &+& \sum_F g_F \left[ {m^2_F\, T^2\over 48} 
- {m_F^4\over 64\pi^2} \left(\ln {m_F^2\over T^2} -c_F \right) \right]
\nonumber \\
&+& \sum_B g_B \left[ {m^2_B\, T^2\over 24} - 
	{m^3_B\, T\over 12 \pi} 
+ {m_B^4\over 64\pi^2} \left(\ln {m_B^2\over T^2} -c_B \right) \right] \,.
\label{eq:high-T-expansions}
\eea
The lowest order potential (\ref{VT}) can be improved by resumming diagrams corresponding to insertions of the thermal mass corrections,
\beq
	m^2_i(H^0,S^0) \to m^2_i(H^0,S^0) + \delta m^2_i
\label{tmass}
\eeq
where $\delta m^2_i$ is of the form $T^2$ times coupling constants. The Debye mass matrices can be obtained by computing the thermal self-energies in the high-temperature limit, but they can also be directly inferred from (\ref{VT}).
For a bosonic degree of freedom $\phi_i$ the Debye mass matrix is 
\beq
\delta m^2_{ij} 
= \sum_k {g_k\over 24}\,{\partial m^2_k\over\partial\phi_i\partial\phi_j}\, T^2 
\label{eq:debye}
\eeq
where $k$ runs over the relevant bosons and fermions in the theory,
and $m^2_{k}$ are the field dependent masses given in Appendix~\ref{masses}\footnote{The sum over a given scalar of fermion representation can be written as $\sum_k g_k m^2_k = {\rm Tr}[m m^\dagger]$.}. Fermions do not get a thermal correction.  
The thermally corrected mass $m^2_i + \delta m^2_i$ in (\ref{tmass}) denotes an eigenvalue of the  full mass matrix $m^2_{ij} + \delta m^2_{ij}$.  

Strictly speaking the ring-improvement is a consistent correction only
for the  bosonic zero modes in the high temperature limit $m/T \ll 1$
\cite{Arnold-Espinosa}, correcting the masses only in the cubic terms
in the bosonic expansion in (\ref{eq:high-T-expansions}). However, we
need the effective potential also for large $m/T$, and we need to be
able to smoothly connect the two regimes. In the $m/T\gsim 1$ regime a
cubic term with eigenvalues computed from high-$T$ Debye mass
corrections (\ref{eq:debye}) becomes a poor approximation for the
effective potential.  The situation could be remedied by computing the
Debye masses for arbitrary $m/T$, whence the Debye corretions would be
exponentially suppressed at large $m/T$. This would be a 
lengthy computation but luckily a more convenient and mathematically
equally consistent prescription is available: one implements
(\ref{tmass}) with the high-T expressions (\ref{eq:debye}) directly in
(\ref{VT}) without making any high-temperature expansion
\cite{Parwani}. Both approaches give essentially indistinguishable
results in the high-$T$ limit while the latter prescription also smoothly
connects to the correct (vanishing) thermal correction at $m/T\gg 1$. 

To complete our approximation we also make the Debye correction within
the 1-loop vacuum correction $V_{\rm CW}$. Indeed, it is known that at
1-loop level the $\ln(m^2_i)$ terms cancel between the $V_{\rm CW}$
and the high-$T$ expansion of $V_{\rm T}$ and we ought to preserve
this property in our ring-improved potential. Log-terms corrected in
this way do not induce spurious nonanalytic behaviour associated
with the negative values of $m^2_i$, which frequently occur for the
Goldstone modes as well as for the physical Higgs in regions where the
curvature of the potential is negative. Negative $m^2_i$ induce an
imaginary part of the effective potential due to the nonanalytic 
cubic $T (m^2)^{3/2}$ term, signalling an instability of homogenous
zero-modes. Here we are only concerned with the real part of the
potential obtained by replacing the cubic term with $T |m^2|^{3/2}$.
Numerically we use the high-$T$ expansion where $m/T$ is sufficiently
small and smoothly match it onto a numerical fit to an exact thermal
integral for larger $m/T$.

\section{Electroweak phase transition}
\label{ewpt}

We wish to identify models which have a first order electroweak phase transition, and which satisfy the
constraint $v_c/T_c > 1$ for the Higgs VEV and critical temperature when bubbles of the broken phase
nucleate.  This constraint ensures that any baryons created during the phase transition are not
significantly depleted by residual sphaleron interactions inside the bubbles.

\subsection{Finding $T_c$}

To determine the strength of the phase transition, we use the following algorithm.  At sufficiently high temperatures one can usually find, through numerical minimization, a minimum of the potential that is in or close to the symmetric vacuum with vanishing Higgs VEVs.  (We discuss exceptions to this statement below).  At low temperatures one can of course find the broken minimum with $\langle H^0\rangle\cong v/\sqrt{2}$.  The first step is to try to bracket the critical
temperature $T_c$ of the phase transition by subdividing an initial
interval $(T_1,T_2)$ in the middle, $T_m=\sfrac12(T_1,T_2)$.  If only
the symmetric minimum exists at $T_m$, then we replace $T_1$ by
$T_m$, whereas if only the broken minimum exists, we replace
$T_2$ by $T_m$.  Continuing in this way, we will either reach an
interval over which both local minima simultaneously exist, in which
case we have bracketed $T_c$ for a first order phase transition, or
else the width of the interval becomes smaller than some cutoff
(we take 0.1 GeV), and the transition is deemed to be second order.
Once $T_c$ is bracketed, one can accurately find $T_c$ and $v_c$ by minimizing
the function
$f(T) = |V_{\rm sym} - V_{\rm br}|$, where $V_{\rm sym}$ and
$V_{\rm br}$ are respectively the values of the potential at the
symmetric and broken minima.

\begin{figure}[t]
\centerline{\scalebox{0.6}{\epsfig{file=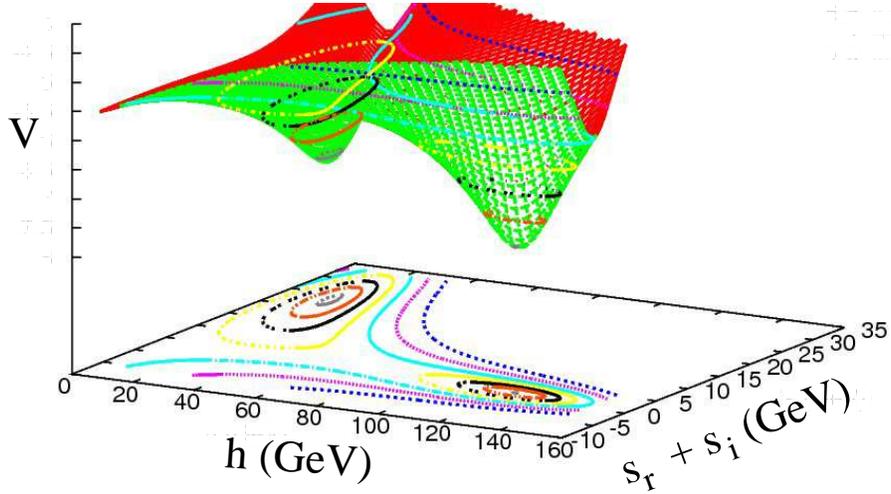}}}
\caption{Potential at $T=T_c$ for a case that exhibits mild
electroweak symmetry breaking in the quasi-symmetric minimum.
The log of [V minus a constant near the minimum value] is plotted
to exaggerate the barrier between the two minima.
}
\label{potfig}
\end{figure}

We find that for nonzero values of $\lambda_4$, the ``symmetric''
minimum in fact sometimes has a sizeable VEV 
$v_s = 
\sqrt{2(|\langle{H^0}\rangle|^2 + |\langle{S^0}\rangle|^2)}$.
As long
as $v_s$ is small, sphalerons in the quasi-symmetric phase are
still fast enough to induce baryogenesis.  However if $v_s$ is
too large, the sphaleron interactions will be suppressed in front of
the bubble wall and lead to a suppressed baryon asymmetry.   We take
the upper bound to be $v_s/T_c < 0.4$.  This is the point at
which the sphaleron rate in the ``symmetric'' phase starts to become
significantly less than its value at $v_s=0$, as can be seen
from Eqn.\ (\ref{fsph}) below.   An example is provided by the parameter set
$m_h=122$ GeV, $m_1 = 185.4$ GeV, $m_2=0$, $\lambda_1 = 1.335$,
$\lambda_2 = 2.553$, $\lambda_3 = -1.291$, $\lambda_4 = -0.114 +
0.434i$, $\lambda_5 = 0$, $\lambda_6=1$, $\eta_t = -0.096$.  The
potential at the critical temperature is shown in Fig.\ \ref{potfig}.
It has $v_s/T_c =0.34$, and so still satisfies our criterion for
sphaleron interactions being fast enough in the symmetric phase.

\subsection{Search strategy}
\label{sssect}

The scalar potential of the general two Higgs doublet model we 
investigate has many parameters.  While it is possible to scan them on
a coarse grid to find many examples satisfying the particle physics 
contstraints discussed in section \ref{constraints}, very few of
these models also have a strong enough phase transition, with $v_c/T_c
> 1$.   This is not
surprising, since imposing constraints lowers the dimensionality of
the allowed parameter space.  For example, searching for points in an
12D space that happen to lie on a complicated 8D subsurface is bound to
be inefficient if we restrict the search to points on a necessarily
coarse grid in the full 12D space.  Such a search could be made more efficient if we had
an analytic formula for the constraint, but this is not possible for
the EWPD.  Table \ref{tab1} shows the results of such grid searches for
several values of $m_h$.  Later we must impose the additional
requirement of getting a large enough baryon asymmetry, which is
difficult to achieve.  It is therefore desirable to maximize the
number of models that satisfy the present constraints.

\begin{table}[t]
\begin{center}
\begin{tabular}{|c|c|c|}
\hline
$m_h$ (GeV) & $\#$ models & $\#$ strong EWPT \\
\hline
115  & 210,000 & 92 \\
120 & 195,000 & 49 \\
130 & 171,000 & 25 \\
\hline
\end{tabular}
\end{center}
\caption{\label{tab1} Results of grid searches of the 2HDM parameter
space, showing the numbers of models that satisfy particle physics
constraints, and of these the number having $v_c/T_c>1$.}
\end{table}

A more efficient search strategy is the Markov Chain Monte Carlo
(MCMC) algorithm.   Instead of blindly searching on a grid, this
method  favors points with a stronger phase transition, while still
attempting to broadly explore the space.  Starting at some
point $\{\lambda_i\}$ in the parameter space, a chain of parameter sets is
accumulated.  A trial step is taken from the current point by varying
the $\lambda_i$ randomly, with step sizes taken from a Gaussian
distribution whose variance can be tuned for optimal performance. 
Let $r$ be the ratio $(v_c/T_c)_{\rm new}/ (v_c/T_c)_{\rm old}$,
comparing the previous point to the new one. The new point is
accepted into the chain if it has $r>1$, and it is accepted with
probability $r$ if $r<1$.  In this way, a chain of models is obtained,
whose probability distribution for $v_c/T_c$ is proportional to 
$v_c/T_c$, thus favoring larger values of $v_c/T_c$.  The occasional
acceptance of points with smaller values of $v_c/T_c$ is helpful 
to avoid getting stuck in a local extremum of the probability
distribution, so that the parameter space is broadly explored.

We modify the basic MCMC procedure by also rejecting the trial point
if it has an unstable vacuum, or if it fails to satisfy any of the
particle physics constraints described in section \ref{constraints}.  We also 
adopt a trick to boost the acceptance rate: the trial point is
alternately generated using the last point or the next-to-last point
in the chain.  Presumably the increase in acceptance results from the
fact that the next-to-last point has already succeeded in leading to
one new point in the chain.  We set the Gaussian widths for the
randomly generated steps to 10\% of the current value for the
parameters $\lambda_1$, $\lambda_2$, $\lambda_3$ and
$\sqrt{|m_1^2|}$, and to $0.1$ for the other dimensionless parameters
$\LfourR$, $\LfourI$, $\lambda_6$, $\zeta_b^0$, $\zeta_t^0$ and the phase $\phi$ of
$\zeta_t^0$.  The width for steps in the mass of the Higgs is 5 GeV.

The above description holds if one is only interested in generating
models with large $v_c/T_c$.  We would also like to favor models
in which the baryon asymmetry can be large.  Therefore we additionally compute
the change of the CP-violating phase of the top quark mass across the
bubble wall ($m_t(z) = |m_t(z)|e^{i\theta(z)}$),
\beq
	\Delta\theta = \theta(z)|^\infty_{-\infty}
\label{deltatheta}
\eeq 
and test the quantity $(\Delta\theta)(v_c/T_c)$ instead of just $v_c/T_c$.
It then becomes consistent, within the philosophy of the MCMC
approach, to impose $v_c/T_c>1$ as another constraint to be satisfied
by all models in the chain.  Similarly, to search for models that
lead to successful baryogenesis as well as the $D\slashed{O}$ dimuon
anomaly, we take $(\Delta\theta)(v_c/T_c) |\mathcal{\zeta}^{'0}_b/(V_{tb}^\star \ V_{ts})|$ as the quantity
to be maximized by the MCMC.

We have noticed that having nonzero $|\lambda_5|$ tends to weaken 
the phase transition.
(Recall that $\lambda_5$ and $m_2^2$ are linked to each
other by our choice of field basis in which $\langle s\rangle = 0$
at  zero temperature, insured by Eqn.\ (\ref{l5rel}).)   This can be understood analytically by considering
a simplified version of the potential,
\beq
	V \sim V_h(h) +\frac12 m_1^2 s^2 +\lambda_5 \, s \, h \, (h^2-v^2)
\eeq
Upon integrating out $s$, it becomes
\beq
	V \sim V_h(h) - {\lambda_5^2\over m_1^2} \,h^2 (h^2-v^2)^2
\eeq
The second term always has positive curvature in the region $h\sim v$,
tending to cancel out any barrier between the symmetric and broken
phases that might arise due to finite temperature effects.  We have
compared MCMC runs in which $\lambda_5$ is fixed at zero to those
where it is allowed to vary and find no significant difference with
respect to the frequency of finding a given value of the baryon
asymmetry.

\subsection{Results}

\begin{figure}[t]
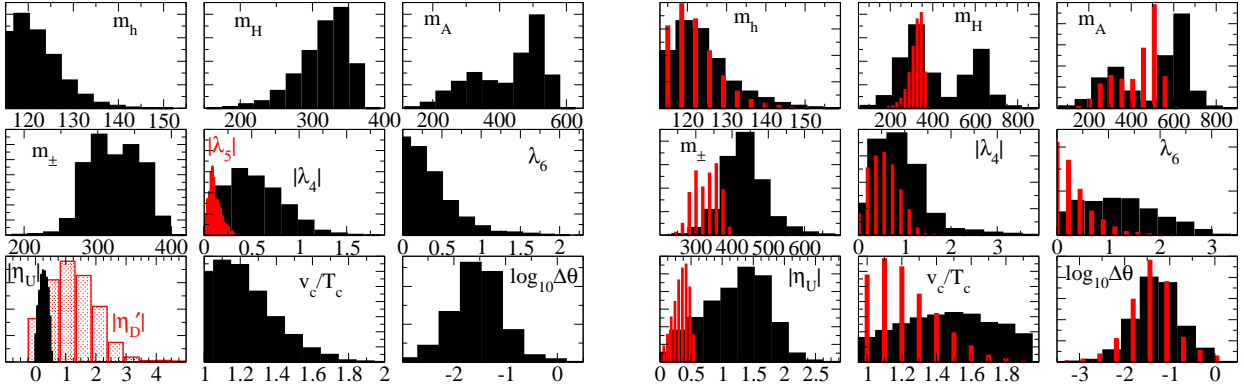

\centerline{\scalebox{0.38}{\epsfig{file=panel4e-etaD.eps}}
\hfill
\scalebox{0.38}{\epsfig{file=panel3-combined.eps}}}
\caption{Left: distributions of parameters satisfying sphaleron and particle
physics bounds, including ${\zeta}^{'0}_b$ and ${\zeta}^{'\pm}_b$, but not insisting on 
reproducing the magnitude of the observed \DO \, 
dimuon excess. $m_{H,A}$ denote masses of the new scalar and pseudoscalar
Higgs bosons, respectively. Normalizations are arbitrary.  Masses are in GeV.
Right: distributions from MCMCs in which ${\zeta}^{'0}_b = {\zeta}^{'\pm}_b \approx 0$, and either omitting
(heavy black bars) or applying (narrow red bars) the constraints from EWPD,
$b\to s \gamma$, neutron EDM, and perturbativity of couplings.
Here $|\eta_U|,|\eta'_D|$ are proxies for 
$|{\zeta}^{0}_t|,|{\zeta}^{'0}_b|$.}
\label{hist1}
\end{figure}

To illustrate some of the statistics of models with a strong phase
transition, we consider a chain of $10^4$ models that pass all of the 
constraints.  In this section and in section VI we use $|\eta'_D|$ as
a more convenient parametrization for ${\zeta}^{'0}_b$ and similarly
$|\eta_U|$ for ${\zeta}^{0}_t$ to avoid having to
rescale by CKM factors and Yukawa couplings in the MCMC.  (The relation
between the $\eta_Q$ and $\zeta_q$ parameters was given in section
 \ref{hot}.)
We do not insist upon the dimuon anomaly here, but we
allow $|\eta'_D|$ to vary so that we can check {\it a posteriori} what
fraction of otherwise allowed models can be compatible with this
potential constraint.  We take the parameter $f=1$ in eq.\ (\ref{generic})
so that the strength of the neutron EDM and $b\to s\gamma$ are 
at a typical level given the assumed value of $\eta'_D$.
Fig.\ \ref{hist1} (left panel) shows the distributions 
of $m_h$,
the new scalar masses $\mH$, $\mA$, $m_\pm$
(where $\mHA$ were defined in 
Eqn.~(\ref{eq:masses-hHA}) and $m_\pm$ are the charged Higgs masses), the dimensionless
couplings $|\lambda_4|$, $|\lambda_5|$, $\lambda_6$, $|\eta'_D|$ 
and $|\eta_U|$, and the quantities
to be maximized (in addition to $|\eta'_D|$): $\Delta\theta$ (Eqn.\ (\ref{deltatheta})) and
$v_c/T_c$.   One sees that relatively low Higgs masses $m_h< 140$ GeV
are necessary,  the masses of the new scalars are less than 350$-$500
GeV for $\SR$ with mass $M_H$ and the charged scalars (the
pseudoscalar $s_I$ with mass $M_A$ tends to
be heavier, $m_A \lsim 500$ GeV), and dimensionless couplings are
typically less than 1 in magnitude with the exception of $|\eta'_D|$,
which we have deliberately pushed toward larger values in the MCMC. 
This parameter has no direct impact on the phase transition
dynamics (since we do not include the $b$ quark contributions to
$V_{\rm eff}$), but it does play a role in
the $R_b$, $b\to s\gamma$, EDM and RGE constraints.

To appreciate the impact of the new particle physics constraints we
have imposed relative to previous studies, we also generated a chain
of models that satisfy only the $v_c/T_c>1$ requirement and the
accelerator bounds on masses.  The resulting parameter distributions
are shown in the right panel of Fig.\ \ref{hist1}.  They are much
broader than those which incorporate the constraints.  This explains in part why we find it more difficult to achieve enough
baryogenesis in the 2HDM compared to earlier investigations. But there are
other important reasons having to do with the production mechanism,
as we describe in the next section.

\section{Baryogenesis}
\label{bau}

To find the baryon asymmetry for a given model, once it has been
established to give a first order phase transition and $T_c$ has been
determined, several steps must be taken.  (1) The solution $\{h(z),
\SR(z), \SI(z)\}$ for the fields in the bubble wall must be
constructed.  We will refer to it as the path taken by the wall in
field space.  (2) A set of coupled Boltzmann transport equations must
be solved for the chemical potentials $\mu_i(z)$ and velocity
perturbations $u_i(z)$ of the particle species considered to be most
relevant for creating a local CP asymmetry in the vicinity of the
wall.  (3) Some linear combination of the $\mu_i$ gives the chemical
potential for left-handed baryon number, $\mu_{B_L}$, which biases
sphaleron reactions in the symmetric phase in front of the wall.
The integral of $\mu_{B_L}$ in the symmetric phase region gives the
baryon asymmetry, up to some proportionality factor.

\begin{figure}[t]
\centerline{\scalebox{0.45}{\epsfig{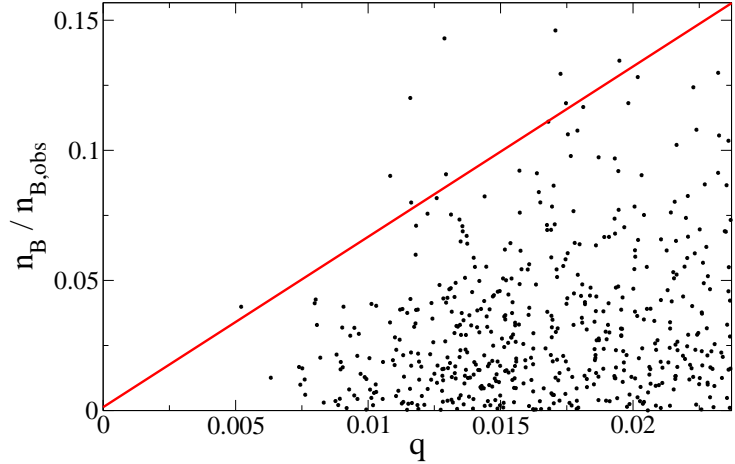}}}
\caption{Scatter plot of baryon number in units of the observed
density ($n_B/n_{B,\rm obs}$) versus the predictor $q$, 
Eqn.\ (\ref{qeq}).}
\label{corr1}
\end{figure}

Our procedure is to carry out these steps for each model in the chains
produced by the MCMC.  We recall that these are models passing all the
phenomenological constraints as well as the sphaleron bound
$v_c/T_c>1$, and also tending to have a large phase change
$\Delta\theta$, and optionally large values of ${\zeta}^{'0}_b$.
We weed out the models that necessarily give
a small baryon asymmetry using a predictor.  
Statistically  the dimensionless quantity
\beq
	q \equiv {v_c\over T_c} \, {\Delta\theta\over L_w T_c}
\label{qeq}
\eeq
where $L_w$ is the bubble wall thickness, tends to be proportional to
the maximum value of the baryon asymmetry that can be produced.  The
correlation is shown in Fig.\ \ref{corr1}.  There are several reasons
that the actual baryon asymmetry can fall below the maximum value
predicted by $q$ (as we will discuss below), but large values of $n_B$ almost always require that
$q \gsim 0.15\, (n_B/n_{B,\rm obs})$.  Therefore if one wants to find the
instances in a chain of 10,000 models that give the largest baryon
asymmetry, it is not necessary to compute $n_B$ for all of them, but
rather focus on those with the largest $q$ values.  Of course, $q$ is
much faster to compute than is $n_B$.

\subsection{Bubble wall path and the CP-violating phase of the top quark}

In this section we discuss our procedure for finding the profiles
of the fields $h$, $\SR$, $\SI$ in the bubble walls, needed for
computing baryon production.  In principle, this task is complicated
because 
 the true bubble nucleation temperature $T_{\rm nuc}$ is
somewhat lower than the  critical temperature $T_c$. A complete
account of the phase transition characteristics would involve finding
$T_{\rm nuc}$, computing the latent heat that drives the transition
and evaluating the frictional forces exerted by the particles and
fields on the expanding wall. Such a calculation would yield both the
wall velocity and the Higgs field profiles over the transition region,
which is what we need for a baryogenesis calculation. This would be a
difficult task however, and its results would still be uncertain
because, {\em e.g.,} reliable techniques to compute the interactions of
the infrared gauge-field modes with the bubble wall do not exist. 
However, generically $T_c-T_{\rm nuc}Ê\ll T_c$ so that $V_{\rm eff}$
changes little between $T_c$ and $T_{\rm nuc}$ and one might expect
that the true profiles are reasonably well approximated by the
solution interpolating between the degenerate minima at $T=T_c$. This
is the approximation we shall adopt here, and the wall veolocity $v_w$
is left as an external free parameter.   The wall profiles are then 
found by numerically solving the equations of motion
\beq
	{\partial^2\phi_i\over\partial z^2} = -{\partial V_{\rm eff}\over \partial \phi_i}
\eeq
at the critical temperature, where $z$ is the direction transverse to
the wall, subject to the boundary conditions that the fields approach
the broken or symmetric minima as $z\to\pm\infty$.

\subsubsection{Gauge dependence of bubble wall profiles}
The coupling between the neutral scalars and the $Z$-field poses a
complication in solving for the bubble wall profiles.
This is because the covariant derivative terms $|D_z \phi_i|^2$
provide a source for the 
classical $Z$-field, and so one is faced by the problem of minimizing
the Hamiltonian:
\beq
\int {\rm d}z\left(|D_z H^0|^2 + |D_z S^0|^2 + V_{\rm eff}(H^0,S^0,T_c) + \frac{1}{2}M_Z^2 Z_\mu Z^\mu + ... \right)
\label{eq:fullham}
\eeq
where $D_z = \partial_z - i(g/2{\rm cos}\theta_W) Z_z$ and dots indicate other $Z$-dependent terms.  Writing $\sqrt{2} H^0 \equiv he^{i\varphi_h}$ and $\sqrt{2} S^0 \equiv se^{i\varphi_s}$ and using the fact that the effective potential depends only on the phase difference $\varphi_h-\varphi_s$ one can show that 
\beq
	\partial_\mu \left( j^\mu_Z - M_Z^2 Z^\mu \right) = 0 \,,
\label{Zcurrent}
\eeq
where $M_Z^2 = g^2(h^2+s^2)/(4\cos\theta_W^2)$ and $j^\mu_Z$ is the current sourcing the $Z$-field: 
\beq
	j^\mu_Z \equiv {g\over 2\cos\theta_W}\left(h^2\partial^\mu\varphi_h + 
	s^2\partial^\mu \varphi_s\right) \,.
\label{Zcurrent}
\eeq
Since all classical fields should vanish far away from the wall, we find that %
\beq
  Z^z(z) = {j^z_Z\over M_Z^2(z)} \,,
\label{classicalZ}
\eeq
while all other components vanish. Recall that in our formulation of
the effective potential, we have chosen a
gauge where the Goldstone boson vanishes, hence $\varphi_h \equiv 0$.
This leads to a 
nonvanishing source (\ref{Zcurrent}) for the $Z$-field since
$\varphi_s$ is generally nonzero. Of course
$Z$-field is gauge-dependent, and by an alternative choice of the
basis for phases \cite{HJLS} we could make $j_Z^\mu \equiv 0$, and in
this gauge $Z^\mu\equiv 0$, by Eqn.~(\ref{classicalZ}).  The price to
pay for this simplicity is that one would have to keep track of four
fields in the effective potential rather than three.

\subsubsection{Spatially varying phase of $m_t$}

A key quantity for computing the baryon asymmetry in the current
scenario is the spatially varying CP-violating phase of the top quark
mass, given {\it a priori} by
\beq
m_t(z) = {y_t\over \sqrt{2}}e^{i\varphi_h} \left(h  + \frac{\zeta_t^0}{y_t} \,  s e^{i\varphi_{sh}}\right)\,,
\eeq
where $\varphi_{sh} \equiv \varphi_s-\varphi_h$. As we have observed, because
of the U(1)$_Y$ gauge symmetry,  $\varphi_h$ and $\varphi_s$ are not both
physically meaningful quantities,  so it is interesting to see how
both gauges discussed above lead to  the same phase for the top quark
mass (especially since different authors have chosen different gauges
in previous treatments). First, in the $Z\equiv 0$  gauge we can solve for the phase
$\varphi_h$ using the constraint $j_Z^\mu=0$:
\beq
\partial_z \varphi_h = - \frac{s^2}{h^2+s^2} \partial_z \varphi_{sh} \,.
\label{eq:hphase}
\eeq
$\varphi_h$ can be reconstructed after the fields $h,s$ and $\varphi_{sh}$ 
are determined using equations following from the reduced Hamiltonian:
\beq
 \int {\rm d}z \left( \frac{1}{2}(\partial_z h)^2 + \frac{1}{2}(\partial_z s)^2
 + \frac{1}{2}\frac{s^2h^2}{s^2+ h^2} (\partial_z \varphi_{sh})^2 
 + \tilde V_{\rm eff}(h,s,\varphi_{sh},T_c)\right) \,,
\label{eq:reducedhamilton}
\eeq
where $\tilde V(h,s,\varphi_{sh})= V_{\rm eff}(h,\SR=s\cos
\theta_{sh},\SI=s\sin \theta_{sh},T_c)$. (We can always shift
$\varphi_h$ into the phase of $S$ in $V_{\rm eff}$ because it
is invariant under global SU(2)$\times$U(1)$_y$ transformations,
and contains no derivatives of the fields.)

Alternatively, if one sets $\varphi_h\equiv 0$, it is necessary to deal
with the nonvanishing $Z_\mu$-field:
\beq
  Z_z(z) = \frac{g}{2\cos\theta_WM_Z^2} s^2 \partial_z \varphi_s \,.
\label{eq:Zfield}
\eeq
This field induces an additonal CP-violating force acting on the top
quark due to the gauge interaction $(g/2\cos\theta_W)\, \bar t
\slashed{Z} \gamma_5 t$.  This term can be removed by a local axial
transformation $t \rightarrow e^{i\gamma^5\varphi_Z/2}t$, which however
reintroduces an additional overall phase $\varphi_Z$ into $m_t$:
\beq
m_t(z) \rightarrow {y_t\over \sqrt{2}}
                   \,e^{i\varphi_Z}\, \left(h  +   \frac{\zeta_t^0}{y_t} s e^{i\varphi_s}\right) \,.
\eeq
It is easy to see that this transformation removes the interaction
term with $Z_z$ given by Eqn.~(\ref{eq:Zfield}) only if $
\partial_z\varphi_Z = - {s^2\over h^2+s^2} \partial_z \varphi_s$ which is
identical to Eqn.~(\ref{eq:hphase}) in the gauge $\varphi_h=0$.
Denoting $\zeta_t^0/y_t = |\zeta_t^0/y_t|e^{i\varphi_\eta}$, we find that the total
phase of $m_t$ is given by
\beq
	\theta(z) = \varphi_h(z) + {\rm
   tan}^{-1}\left(| \frac{\zeta_t^0}{y_t}|s\sin(\varphi_\eta + \varphi_{sh})\over
	h + | \frac{\zeta_t^0}{y_t}|s\cos(\varphi_\eta + \varphi_{sh})\right) \,.
\label{thetaeq}
\eeq
It is noteworthy that both contributions to the overall phase are
suppressed by powers of $s/h$: there is no phase if the bubble wall
profile remains strictly along the $h$ direction. Bending of the path
in the $s$ directions is necessary for generating any CP violation in
the wall.  

Although it might appear from (\ref{eq:hphase}) that one more
integration is required to solve for $\theta(z)$, in fact we never
need $\theta(z)$ itself, but only $\partial_z\theta$, to evaluate
the CP-violating source term of the Boltzmann equations, to be
discussed below.  Eqn.\ (\ref{thetaeq}) thus constitutes a sufficient
solution for the phase of $m_t$ in the bubble wall.

\begin{figure}[t]
\centerline{\scalebox{0.92}{\epsfig{file=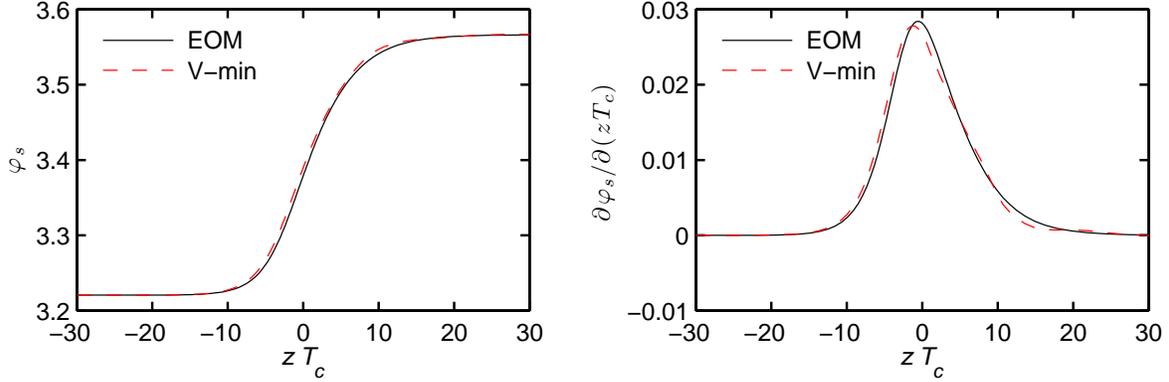}}}
\caption{Left: the phase of $S^0$-field computed from the full EOM (solid black) and from the 
minimization of $V$ (red dashed) as a function of $zT$ corresponding 
to  model 1 in table \ref{tab:largeetaDmodels}.  
Right: same for the $zT$ derivative of the phase.}
\label{fig:compare}
\end{figure}

\subsubsection{Numerical solution for bubble wall profiles}

The equations of motion can be solved numerically by shooting or by
relaxation. However, for the purposes of doing an MCMC search, a
simpler and faster approximate method is desirable.  We first find the
straight line in field space that connects the two VEVs and then 
locate, for each point on this line, the position along the plane
perpendicular to the line where the potential is minimized.  This
deforms the line into a curve, which is
the path of steepest descent through the saddle point. 
One then solves the constrained one-dimensional problem along
this path in the field space to determine the profile. This procedure is valid if the path
does not curve very much (which we typically find to be
true), but it tends to allow the path to curve more than the actual
solution, since it ignores gradients of the  fields when it determines
the path. The  real solution is systematically smoother than than the
approximate one, which tends to overestimate the baryon asymmetry by a
factor of $\lsim 2$. The approximate method is thus sufficient for
searching good candidates in our MCMC-runs, but for accurate results 
we recompute profiles for all models that pass the constraints by
solving the equations of motion by use of accurate relaxation
methods.

\begin{figure}[t]
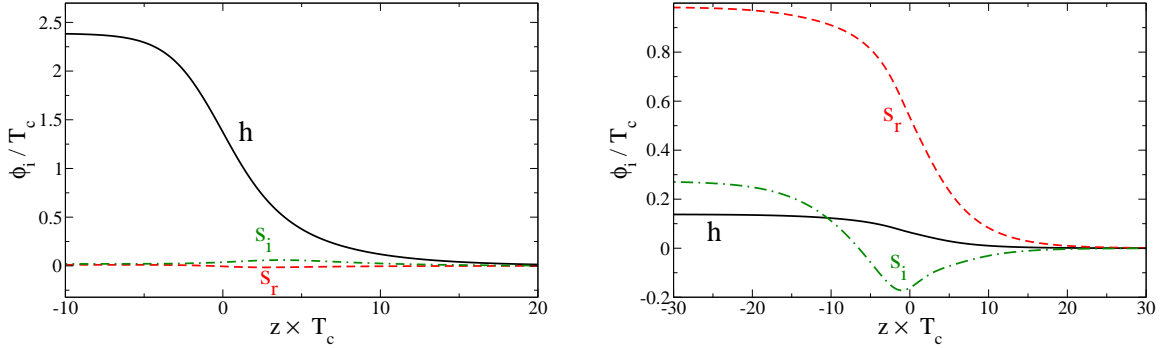

\centerline{\scalebox{0.33}{\epsfig{file=profile-example.eps}}
\hskip0.7truecm
\scalebox{0.33}{\epsfig{file=profile-example-big.eps}}}
\caption{Examples of bubble wall profiles for the fields
$h$, $\SR$, $\SI$.  Left: a typical profile; right: a profile leading
to an atypically large CP-violating phase $\theta(z)$, and using
model parameters that are ruled out by $R_b$ and $b\to s \gamma$
constraints. }
\label{profile1}
\end{figure}

\begin{figure}[t]
\centerline{\scalebox{0.38}{\epsfig{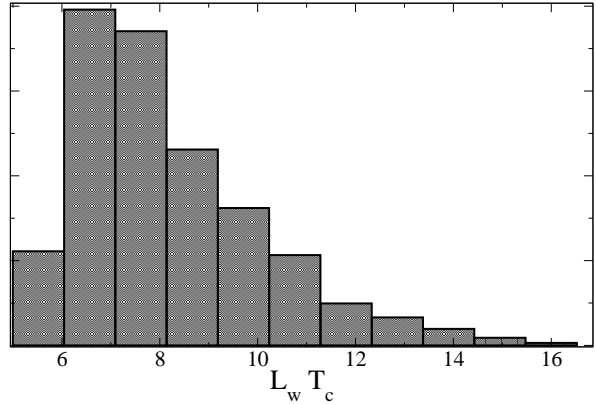}}}
\caption{Distribution of bubble wall thicknesses $L_w$, in units of
$T_c^{-1}$  }
\label{histlw}
\end{figure}

In Fig.~\ref{fig:compare} we show the solutions for the phase
$\varphi_{sh}$ (left panel) and it derivative (right panel) as a function
of $zT_c$ using the minimization of the potential (dashed curves) and
the full equations of motion (solid curves) for the second model
listed in table \ref{tab:largeetaDmodels}. As claimed, the true
solution is smoother than the one determined by constrained
minimization of the potential. In Fig.\ \ref{profile1} we display the profiles
$\{h,\SR,\SI\}$ as a function of $zT_c$ for some other typical models.
Based on these figures the wall thickness appears to be of order
$10/T$. This is an important parameter for electroweak baryogenesis
and although we do not approximate the wall profiles by any analytic
ansatz, it is convenient to characterize the wall thickness $L_w$ by
fitting the total VEV $v(z)$ to the form $\sfrac12 v_c
\tanh(z/L_w)$. In this way we find that $L_wT_c \sim 6-12$ for most
models in our chains. The distribution of $L_w$ values is shown in
Fig.\ \ref{histlw}. 

\subsubsection{Comparison with previous results}

We have noted that the CP-violating phase $\theta(z)$ of the top quark mass
is suppressed by the ratio $s/h$ in the bubble wall, and that
$s/h$ is typically small for models that are not ruled out by 
particle physics constraints (see Fig.\ \ref{profile1}).  This is 
in contrast to the earlier analysis of 
ref.\ \cite{FHS}.  Contrary to this one and earlier papers, we are finding a
systematic suppression in the magnitude of $\theta(z)$	that goes
beyond the smallness of CP-violating phases in the potential.  Namely,
since $\langle S\rangle = 0$ at zero temperature, but $\theta(z)$
vanishes if $S$ were to remain zero throughout the bubble wall, any
nonvanishing $\theta(z)$ comes about as a finite-temperature effect,
{\it i.e.,} $\langle S\rangle$ being no longer zero in the broken minimum
at finite $T$.  On the other hand, ref.\ \cite{FHS} did not observe
any such suppression.  The reason is that ref.\ \cite{FHS} did not
actually solve for the bubble wall profiles, but made the assumption
that the two Higgs fields remained equal to each other in the wall,
$|H_1|=|H_2|$ (which would translate into $|S|=0$ in our field basis).
This assumption is only true if $H_1\leftrightarrow H_2$ is a symmetry
of the Lagrangian, which ref.\ \cite{FHS} claimed to be the case due to a choice of
relations between the parameters of the Higgs potential.  It would
require $\lambda_4 = \lambda_5 = 0$ in our language.  But it is clear
in that case that 
$|S|$ remains strictly zero in the bubble wall, so that
$\theta(z)=0$.  In the language of ref.\ \cite{FHS}, the CP-violating
phases in the Higgs potential should vanish in order to have the
symmetry  $H_1\leftrightarrow H_2$ leading to $|H_1|=|H_2|$ in the 
wall, but this would also have made $\theta(z)=0$.  In short, ref.\
\cite{FHS} made inconsistent assumptions which obscured the suppression of $\theta(z)$ that we have made explicit.

\subsection{Transport equations and source term}

The next step in determining the baryon asymmetry is to solve for the chemical potentials of relevant
particle species in the vicinity of the bubble wall, that are induced
by the top quark mass phase $\theta(z)$.  One approach to deriving
these equations is to start with the Boltzmann equations and to
perturb the particle distribution for species $i$ away from its
equilibrium form by allowing for a small chemical potential $\mu_i$
and a perturbation $\delta f_i$ that describes the departure from
kinetic equilibrium \cite{CJK}:
\beq
	f_i = {1\over \exp(-\beta\gamma_w(E +v_w p_z)- \mu_i) \pm 1} + \delta f_i
\eeq
where $v_w$ is the bubble wall velocity and $\gamma_w = (1-v_w^2)^{-1/2}$.  
Linearizing in $\mu_i$ and $u_i = \langle \delta f_i v_z\rangle$,
the perturbation to the particle's fluid velocity, leads to a set of
coupled first order transport equations for $\mu_i$ and $u_i$
 \cite{CJK}.  This procedure has been carried out for the 
2HDM in refs.\ \cite{FHS,FH} and we adopt the same transport equations as in 
\cite{FHS} which were tailored to two Higgs doublet models. Here the
CP-violating phase of $m_t$ provides the source term and the species
that are followed are the quarks $t_L$, $t_R$, $b_L$ and the Higgs
bosons, which are assumed to have a common chemical potential $\mu_h$
and velocity perturbation $u_h$.

The transport equations depend upon the bubble wall velocity $v_w$,
which we take to be an undetermined parameter.  In the computations
that follow, we have chosen $v_w = 0.1$ for definiteness, but we have
checked that the final results for the baryon asymmetry do not 
depend strongly on this choice over a range of reasonable 
values.  In fact the most important $v_w$-dependence is that which multiplies
the source term, which we discuss next.  However, in the end this 
dependence gets canceled by an explicit factor of $1/v_w$ in the 
relation between the chemical potentials and the baryon asymmetry,
eq.\ (\ref{etabint}) below.  The residual dependence of the result
on $v_w$ comes only through its appearance in subleading terms in the
transport equations, which explains the relatively weak effect.  For
example we find that the baryon asymmetry increases by only 20\% 
when taking $v_w \to 0.05$.  (Ref.\ \cite{Tulin:2011wi} which uses
a different form for the diffusion equations finds a stronger
dependence.)

\subsubsection{The source term}

\begin{figure}[t]
\centerline{\scalebox{0.92}{\epsfig{file=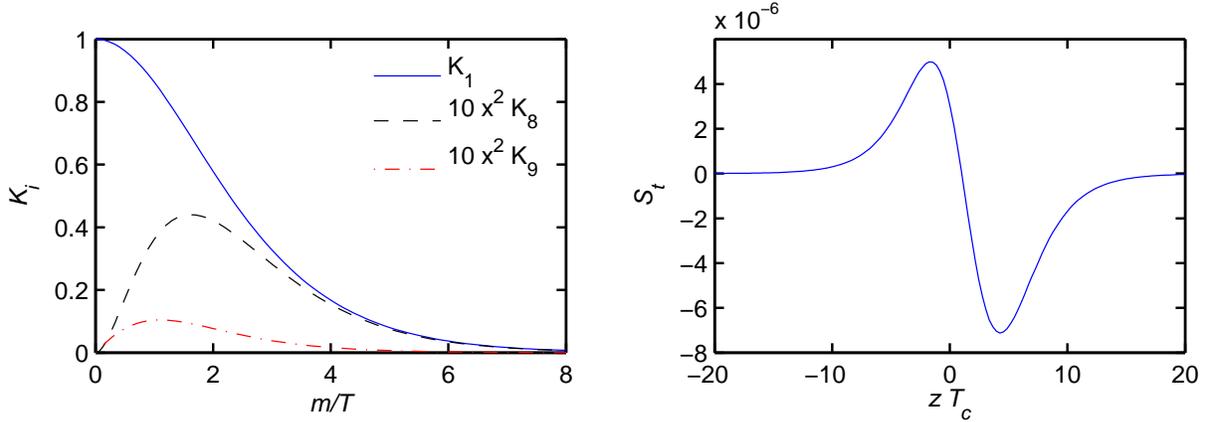}}}
\caption{Left: thermal functions $K_{8,9}$ appearing in the top quark source term, Eqn.\ (\ref{sourceterm}), and $K_1$ which appears in the $B_L$ chemical potential, Eqn.\ (\ref{mubleq}).  
Right: an example (model 1 in table \ref{tab:largeetaDmodels}) of 
the source term as a function of $zT$.  In this case the baryon 
asymmetry computed from full and 1D approximation to the field 
equations agree to 1\%. }
\label{Kfns}
\end{figure}

 A crucial ingredient in the transport equations is the
inhomogeneous source term, which after many years is still a matter of
controversy in the EWBG community.  The source term adopted by
\cite{FHS} and by us encodes the fact that particles, and here the
top quarks in particular, experience a CP-violating  force while
traversing the bubble wall, where $\theta(z)$ is  varying
\cite{JPT,CJK,CK,KPSW,PSWI}. The corresponding force term in the
Boltzmann equations thus gives rise to the (scaled dimensionless)
source 
\beq
S_t = v_w \left( 
    K_8(x_t) (x^2_t\theta')' -  K_9(x_t)x_t^2 x_t^{2\prime}\theta' 
          \right) \,,
\label{sourceterm}
\eeq
where primes denote $\partial_{zT}$ and $K_{8,9}$ are dimensionless
functions of $x_t \equiv |m_t|/T$ arising from phase-space averaging
of certain kinematic variables,\footnote{Our $K_{8,9}$ are related to
those of \cite{FHS} by simple $T$ scalings that make them
dimensionless.} plotted in Fig.\ \ref{Kfns}.

Since the typical wall thickness is significantly larger than
$T_c^{-1}$, the derivatives in $S_t$ are a source of suppression of
the baryon asymmetry.  A source term with fewer derivatives could
possibly give larger results.  Such contributions
arise from corrections due to CP-violating dispersion relations to
 collision integrals in the Boltzmann equations. For example the 
Higgs decay to $t\bar t$ pairs contributes a source 
\cite{PSWII}:
\beq
S_\phi = - v_w {y_t^2\over 8\pi^3 }\, I_\phi\, \Theta(M_H-2|m_t|) 
x_t^2\theta' \,,
\label{eq:coll-source}
\eeq
where $M_H$ is the mass of a Higgs boson, $\Theta$ is the Heaviside function and $I_\phi$ is a thermal kinematic function of $(M_H/T, |m_t|/T)$ typically of
order $\sim 0.1$.  However because of its small coefficient we find that the
contribution of $S_\phi$ to the baryon asymmetry is always at least 10
times smaller than that of $S_t$.

 The controversy alluded to above stems from the fact
that there are competing formalisms for deriving source terms for the
transport equations, that originate from refs.\ \cite{Riotto,CQRVW}.
These attempt to capture the quantum mechanical aspects of the
particle interactions with the wall, rather than the semiclassical
behavior described by the sources
(\ref{sourceterm}-\ref{eq:coll-source}).  A difficulty with this
approach is that rather drastic approximations need to be made in
order obtain concrete, tractable expressions, notably expanding
propagators to leading order in VEVs of the spatially varying Higgs
fields (the ``VEV insertion approximation''). The resulting source
term has only a single derivative, unlike the semiclassical force
source term we have adopted.  It thus gives a larger estimate for the
baryon asymmetry. However we do not believe that it represents a
controlled approximation, so we adhere to the semiclassical approach.
A more complete transport formalism, capable of
accounting for the nonlocal coherence effects associated with the
quantum interactions, is being developed in refs.~\cite{HKR}.
There has been recent progress in improving upon the VEV insertion
approach in \cite{Cirigliano:2009yt,Cirigliano:2011di}, although
so far only with source involving scalar fields.

\subsubsection{Transport equations}

To solve the transport equations of ref.\ \cite{FHS} accurately, it is necessary to use
the relaxation method, since the asymptotic behavior of the solutions
in the large $|z|$ regions cannot be determined analytically.  On the
other hand, shooting from large values of $\pm |z|$ toward $z=0$
is computationally much more efficient.  The problem with shooting
is that in principle one needs to know the analytic behavior of the
asymptotic solutions in order to correctly specify the ratios 
$\mu_i/u_i$, since only half of the independent variables can be input
as free parameters (to be determined by smoothly matching the
solutions at $z=0$) at either boundary.  For example, one could freely
vary $u_i$ at each boundary while holding $\mu_i$ fixed to find a
solution, but one does not know beforehand what values of $\mu_i$ to
choose.  If the starting points are sufficiently far from $z=0$, then
it might be a good approximation to take $\mu_i=0$ at the boundaries.
But since the solutions are exponentially decaying functions of $|z|$,
shooting is not numerically stable if the boundaries are taken to be
so distant that setting $\mu_i=0$ would be a good approximation.  

\begin{figure}[t]
\centerline{\scalebox{0.92}{\epsfig{file=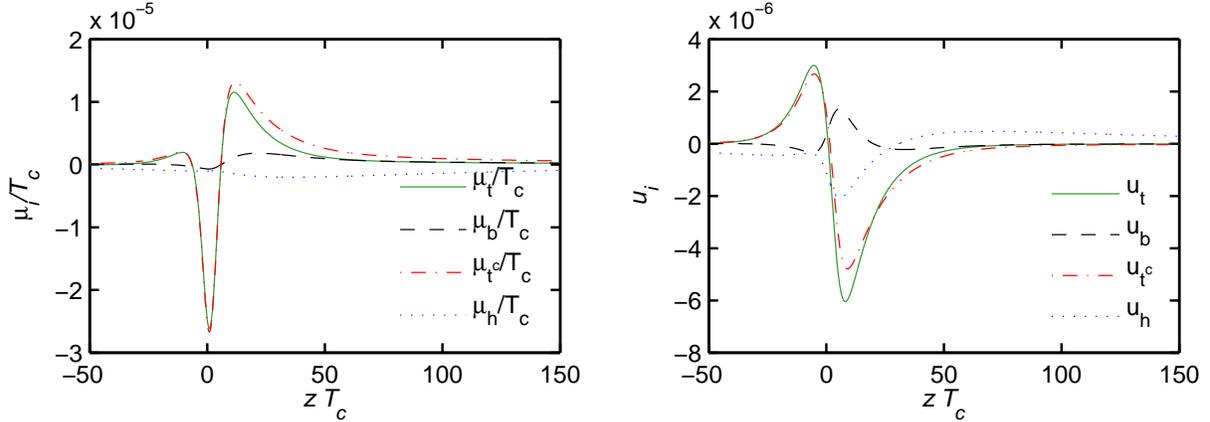}}}
\caption{Typical example of solution to transport equations for the
chemical potentials of $t_L$, $t_L^c$, $b_L$ and Higgs bosons.
This example (model 1 from table \ref{tab:largeetaDmodels})
gives a baryon asymmetry which is $1.1$ times that of the
observed value.}
\label{mufig}
\end{figure}

Of course the extra computational burden of relaxation is not an issue
if one only needs to solve the transport equations for a small number
of models, but to check thousands of models generated by the MCMC, a
compromise is needed.  We find that shooting typically gives an
estimate for the baryon asymmetry that is accurate to within a factor
of a few.  We reserve relaxation for refining such estimates for
potentially interesting examples.  Typical examples of
solutions for $\mu_i(z)/T_c$ and $u_i(z)$ from the relaxation method
calculation are shown in Fig.\ \ref{mufig}.

%
\subsection{Baryon asymmetry results}
%

\begin{figure}[t]
\centerline{\scalebox{0.55}{\epsfig{file=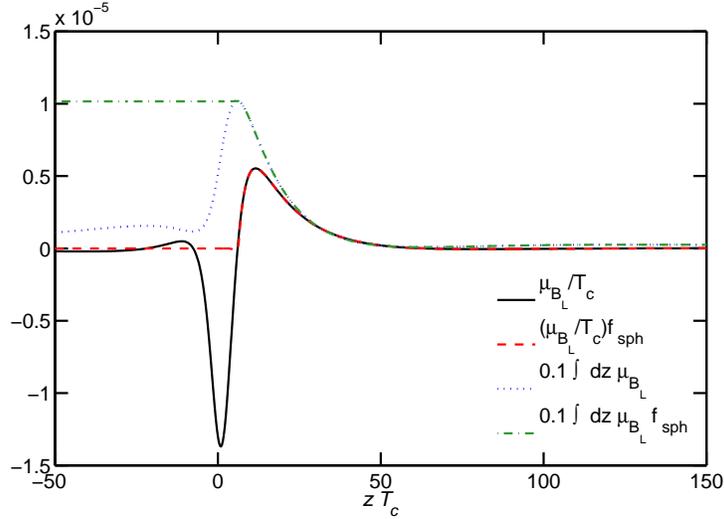}}}
\caption{Example of $z$-dependence of $\mu_{B_L}$, $\mu_{B_L} f_{\rm sph}$
and their integrals, showing the importance of the sphaleron rate
modulation factor $f_{\rm sph}$.  Symmetric phase is on the right.}
\label{mublfig}
\end{figure}

Once the chemical potentials around the bubble wall are known, they
determine the overall asymmetry in left-handed baryon number,
$\mu_{B_L}$, which biases the sphaleron interactions.  Following
\cite{FHS}, we take this to be 
\beq
	\mu_{B_L} = \sfrac12(1 + 4K_{1,t})\mu_t + 
	\sfrac12(1 + 4K_{1,b})\mu_{b} - 2 K_{1,t}\mu_{t^c}
\label{mubleq}
\eeq
where $K_{1,i}$ is a dimensionless function of $m_i/T$ shown in Fig.\ \ref{Kfns}.  
(Notice that $K_{1,b}\cong 1$ since $b$ is nearly massless.) The baryon-to-entropy ratio can then be 
calculated as
\beq
	\eta_B = {405\,\Gamma_{\rm sph}\over4\pi^2 v_wg_* T}
	\int dz\, \mu_{B_L}\, f_{\rm sph} \, e^{-45\,\Gamma_{\rm sph}|z|/(4 v_w)}
\label{etabint}
\eeq
where $\Gamma_{\rm sph}\approx 10^{-6} T_c$ is the rate of sphaleron
interactions in the symmetric phase, the exponential factor takes care
of washout in the symmetric phase in case $v_w \ll 1$,  and $f_{\rm
sph}$ takes into account the $z$-dependence of the actual sphaleron
rate as it turns off in the broken phase.  We estimate
$f_{\rm sph}$ as follows: in the broken phase where the Higgs VEV
$v = \sqrt{2(|H^0|^2 + |S^0|^2)}$ is nonzero, the sphaleron rate is
given by\footnote{To be more precise the factor 40 in the exponent
should be replaced by a function depending on the couplings in the
scalar sector, but that would not change our results essentially.}
$\Gamma_{\rm bp} \sim Ne^{-40 v/T}$. We fix $N$ by equating 
$\Gamma_{\rm bp}\equiv H_c^{-1}$ at $v/T_c=1$ (this encodes the usual
sphaleron wash-out constraint $v/T_c > 1$), giving $\Gamma_{\rm bp} =
{H_c}^{-1} e^{40(1-v/T_c)} \approx 2.4 \;T_c \; e^{-40v/T_c}$, where
we used $H_c\approx 10^{-17}T_c$. We now assume that this
equation holds also for $v = v(z)$ within the wall, and define the
overall sphaleron rate to be ${\rm min}(\Gamma_{\rm sph},\Gamma_{\rm
bp}(z)) \equiv \Gamma_{\rm sph} f_{\rm sph}(z)$ which 
gives 
\beq
	f_{\rm sph}(z) = {\rm min}\left(1,\, {2.4\, T\over 
	\Gamma_{\rm sph} }\,\mathit e^{-40\, v(z)/T}\right) \,.
\label{fsph}
\eeq 
It is important to use a realistic estimate for $f_{\rm sph}$ rather
than a simple step function, because we find that the integral of
$\mu_{B_L}$ over the symmetric phase can be much smaller than that of
$\mu_{B_L} f_{\rm sph}$. This method also treats correctly those cases
(and automatically cuts out the most extreme) where the VEV is nonzero even in
the symmetric phase. The effect of the $f_{\rm sph}$ modulation is
illustrated in Fig.\ \ref{mublfig}.

\begin{figure}[t]
\centerline{\scalebox{0.45}{\epsfig{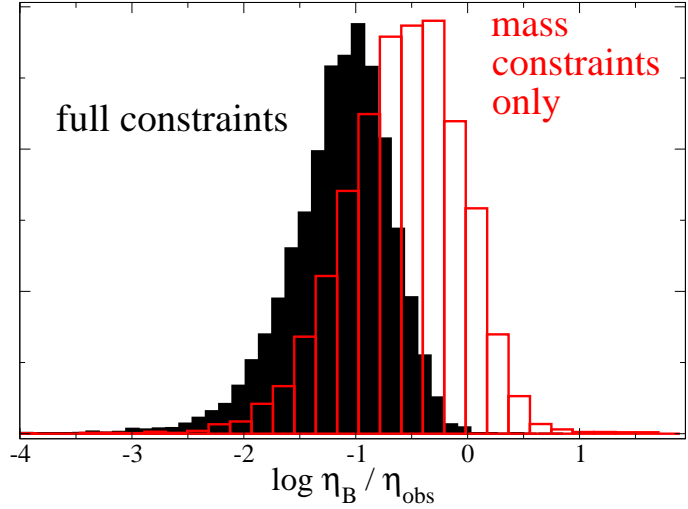}}}
\caption{Solid bars: frequency of $\eta_B$ values from chain of $10^4$ models that
satisfy all constraints.  Open (red) bars: frequency of $\eta_B$ when
the $R_b$, $b\to s\gamma$ neutron EDM and perturbativity constraints are omitted. }
\label{baufig}
\end{figure}

We have searched the parameter space for models likely to provide a large
enough baryon asymmetry and, optionally, a large enough dimuon excess in
addition.  We consider two cases. In the first, we set ${\zeta}^{'0}_b =
{\zeta}^{'\pm}_b = 0$ so that the constraints from $R_b$, $b\to s\gamma$ and the
neutron EDM are relaxed as much as possible without having to fine-tune the
parameter $f$ in eq.\ (\ref{generic}).   In this case we have or course given up
on trying to explain the dimuon excess.  We find that it is difficult to achieve
a large enough baryon asymmetry even without this extra demand.  A chain of 10,000 models  designed to achieve large $v_c/T_c$ and
$\theta(z)$ yields no models that have $\eta_{B}$ significantly greater than the
observed value, and only a handfull which come close to the observed value.  Rarely
we find examples where the Landau pole occurs above the scale $\Lambda = 3$ TeV;
working models with $\Lambda \cong 1$ TeV are much more frequent.    The
distribution of values of $\eta_{B}$ for these 10,000 models is shown in Fig.\
\ref{baufig}.  It is remarkable that the upper limit on the baryon asymmetry
happens to be so close to the observed value.   In the figure we also show the
distribution of $\eta_B$ in a chain where the constraints from EWPD, $R_b$,  $b\to
s\gamma$ and the neutron EDM are omitted; in that case it is easier to reproduce
the observed baryon asymmetry, although there is still relatively little room for
producing too great an asymmetry.  Parameters for a few working models are given
in table \ref{tab2}.  They share several properties: light SM-like Higgs masses
$\sim 117-125$ GeV, new neutral and charged Higgs masses near 300 GeV, and values
of $|{\zeta}^{0}_t|\sim 0.5$ that saturate the constraint coming from $R_b$, with
${\zeta}^{0}_t $ having a  large phase $\theta$.

\begin{table}[t]
\begin{center}
\begin{tabular}{|c|c|c|c|c|c|c|c|c|c|c|c|c|c|c|c|}
\hline
$m_h$& $m_1$ & $m_H$ & $m_A$ & $m_\pm$ & $\lambda_1$ & $\lambda_2$ & $\lambda_3$ & 
 $\LfourR$ & $\LfourI$ &  $\lambda_6$ & $|\eta_U|$ &
 $\theta$  & ${v_c\over T_c}$ & $\Lambda$ & ${\eta_B\over\eta_{\rm obs}}$\\
\hline
 $117$ & $77$  & $289$ & $327$  & $299$ & $2.96$ & $ 0.19$ & $-0.20$ &    $-0.56$ &     $-0.17$ &    $0.04$ & $ 0.42$ &$-0.97$ & $1.50$ & $3304$ & $1.00$ \\
 $121$ & $141$ & $300$ & $407$  & $307$ & $2.46$ & $ 1.12$ & $-0.63$ &    $\phm0.10$  & $\phm0.25$ & $1.96$ & $ 0.52$ & $\phm3.83$ & $1.10$ & $1019 $ & $1.32$ \\ 
 $125$ & $121$ & $374$ & $325$  & $310$ & $2.68$ & $ 0.89$ & $\phm0.28$ & $-0.22$ &     $-0.27$ &    $1.66$ & $ 0.52$ & $\phm3.41$ & $1.07$ & $1146$ & $ 1.20$ \\
 $123$ & $129$ & $263$ & $399$  & $292$ & $2.26$ & $ 0.97$ & $-0.74$ &    $\phm0.12$ &  $\phm0.22$ & $2.14$ & $ 0.43$ & $\phm3.68$ & $1.04$ & $1007$ & $1.18$ \\
\hline
\end{tabular}
\end{center}
\caption{\label{tab2} Parameters of sample models with ${\zeta}^{'0}_b = {\zeta}^{'\pm}_b = 0$
(hence $\eta'_D=0$, and no dimuon anomaly)
and sufficient baryogenesis, and a Landau pole at scale $\Lambda$.
Masses/energies are in GeV.}
\label{tab:smalletaDmodels}
\end{table}  
\begin{table}[t]
\begin{center}
\begin{tabular}{|c|c|c|c|c|c|c|c|c|c|c|c|c|c|c|c|c|c|}
\hline
$m_h$& $m_1$ & $m_H$ & $m_A$ & $m_\pm$ & 
$\lambda_1$ & $\lambda_2$ & $\lambda_3$ & $\LfourR$ & $\LfourI$ &  $\lambda_6$ & $|\eta_U|$ & $\theta$ &  $|\eta'_D|$ & $|\eta'_{\DOO}|$ & ${v_c\over T_c}$ & $\Lambda$ & ${\eta_B\over\eta_{\rm obs}}$\\
\hline
$121$ & $212$ & $350$ & $523$  & $355$ & $2.67$ & $2.39$ & $-1.25$ & $0.630$ & $-0.545$ & $0.08$ & $0.560$ & $-2.25$ & $4.79$ & $4.4$ & $1.65$ & $1675$ & $1.14$ \\  
$117$ & $168$ & $276$ & $453$  & $316$ & $2.36$ & $1.35$ & $-1.06$ & $0.052$ & $-0.215$ & $1.40$ & $0.381$ & $-2.00$ & $5.95$ & $3.3$ & $1.24$ & $1382$ & $1.04$ \\
$119$ & $199$ & $301$ & $485$  & $318$ & $2.03$ & $2.06$ & $-1.20$ & $0.046$ & $-0.168$ & $1.73$ & $0.463$ & $-2.23$ & $4.41$ & $3.7$ & $1.22$ & $1160$ & $1.22$ \\
\hline
\end{tabular}
\end{center}
\caption{\label{tab3} Sample of model parameters giving sufficiently large 
$|\eta'_D|$ (hence $\zeta'^0_b$)
and sufficient baryogenesis.  $\eta'_{\DOO}$ is the value needed to match the
central value of the dimuon anomaly.}
\label{tab:largeetaDmodels}
\end{table}

In the second case, we bias the Monte Carlo to produce models with large values
of $|{\zeta}^{'0}_b|$ to try to simultaneously explain the dimuon excess. 
In this scenario, while it is more difficult to achieve a large enough
baryon asymmetry than in the previous one, the
results are rather sensitive to the choice of $f$ in (\ref{generic}), with $f\to
0$ giving the most  relaxed constraints.   Similarly to the previous case, with a
Landau pole cutoff of 2 TeV, we find no viable examples but at $1$ TeV and we
find several  (out of $10^4$ models in the chain) that have a large enough baryon
asymmetry.  If $f=1$,  the actual values of $\eta'_D\sim 1$-$2.5$ fall below the
central values $3.6$-$4.6$ needed to agree with the \DO\ dimuon observation. 
However a modest relaxation of the constraints by using $f=0.5$ leads to more
favorable results, where we find $\sim 10$ models with the desired baryon
asymmetry and dimuon signal.  (The distribution of models satisfying both constraints 
is displayed graphically in the final figure in section \ref{conclusions}.)
These examples are similar to those with 
${\zeta}^{'0}_b = {\zeta}^{'\pm}_b \approx 0$ in that the SM-like Higgs must be
light, while the new scalars exceed 300 GeV in mass.

\section{Experimental testability}
\label{collider}

We have shown that a direct link between the new phase hinted at in $B$ physics data and  EWBG cannot be
made in 2HDM models in general. EWBG depends upon CP violation in the scalar potential and top quark
Yukawa couplings, while enhanced $B_s$ mixing involves new couplings to the $b$ quark and phenomenological constraints
severely limit the latter couplings.  Nevertheless,
the prospects for direct collider confirmation (or refutation) of the new phase hinted at in the dimuon
data and the new scalar  degrees of freedom we have discussed are very strong. In this section we briefly
discuss some of these prospects.

\subsection{Confirmation/refutation of the dimuon anomaly}

The hint for new physics near the electroweak scale indicated by the \DO\ collaboration
\cite{Abazov:2010hv} that has at least one new source of CP violation,   is consistent with other
direct measurements that are sensitive to the same phase in $B_s$ mixing and use flavour tagged $B_s^0
\rightarrow J/\Psi \, \phi$ decays.  With $6.1 \, {\rm fb^{-1}}$ of data, \DO\ \cite{D0-6098-CONF} finds
a hint of a new phase in decays of this form; their results show a larger deviation from the SM than the
equivalent measurements by CDF  \cite{CDF-10206}. Both results are statistically consistent.
A more recent measurement that has appeared since the original version of this work includes
$8 \, {\rm fb^{-1}}$ of data from \DO\  reported in \cite{Abazov:2011ry}. A method for
 checking the consistency of these
results can be found in \cite{Ligeti:2010ia,Grossman:2009mn}:
one can compare the fit values of $\Delta \Gamma_s$ and $S_{\phi \psi}$ from the two
measurements.  The relevant relation is
\bea
|\Delta \Gamma_s| \simeq - \left[0.20 \pm \, 0.11 {\rm ps}^{-1} \right] \sqrt{1 -S_{\phi \psi}^2}/S_{\phi \psi},
\eea
using $\delta m_s = 17.78 \pm 0.12 \, {\rm ps}^{-1}$, $a_{SL}^d = - (4.7 \pm 4.6) \times 10^{-3}$ and  $a_{SL}^b = - (0.787 \pm 0.172 \pm 0.093) \times 10^{-2}$.
Using $S_{\phi \psi} = -0.55 \pm 0.35$ one finds that \cite{Abazov:2011ry} implies
\bea
\frac{|\Delta \Gamma_s| \, S_{\phi \psi}}{\sqrt{1 -S_{\phi \psi}^2}} \simeq -0.11 \pm 0.10 \, {\rm ps}^{-1}
\eea
which is consistent with the updated value derived from the dimuon anomaly. It is encouraging that \DO \, has just announced that
with $9 \, {\rm fb^{-1}}$ of data the significance of this deviation in the dimuon measurement is now $3.9 \sigma$ compared directly
against the SM prediction of the dimuon charge asymmetry.

The first preliminary measurements of \cite{LHCb-prelim} flavour tagged $B_s^0 \rightarrow J/\Psi \,
\phi$ decays at $\rm LHC_b$  using $36 \, {\rm pb^{-1}}$ of data found a (statistics limited) result that agreed well with the past measurements of
\DO \cite{D0-6098-CONF} and CDF \cite{CDF-10206}, and was consistent with the dimuon
anomaly. However a more precise measurement was subsequently reported at Lepton-Photon 2011 that introduces significant tension into
the interpretation of the pattern of data as implying new physics.
This measurement reports $\Delta \Gamma_s \simeq 0.123 \pm 0.029 \pm 0.008$ and $S_{\phi \psi} =0.13 \pm 0.18 \pm 0.07$,
which gives
\bea
\frac{|\Delta \Gamma_s| \, S_{\phi \psi}}{\sqrt{1 -S_{\phi \psi}^2}} \simeq 0.021 \pm 0.024 \, {\rm ps}^{-1}.
\eea 
These results are to be compared to the result derived from the updated dimuon anomaly data 
$- 0.20 \pm \, 0.11 {\rm ps}^{-1}$ --
demonstrating a tension at about the $2 \sigma$ level.

This new $\rm LHC_b$ result is not a direct measurement of the dimuon asymmetry, but 
the prospect for such a determination is promising. In
part due to the improved mass resolution at $\rm LHC_b$, a measurement of the difference in the asymmetry
in $B_s$ mixing ($a_{fs}^s$) and $B_d$ mixing ($a_{fs}^d$) is being pursued. This approach has the added
benefit of detector asymmetries (but not production asymmetries) cancelling through taking the
difference. For both $a_{fs}^s$ and $a_{fs}^d$, semileptonic $B \rightarrow D \, \mu^- \, \bar{\nu}$
decays and subsequent kaon-rich decays of $D_s^+ \rightarrow K^+ \, K^- \, \pi^+$ and $D^+\rightarrow K^+
\, K^- \, \pi^+$ are used for event selection. Measuring  $a_{fs}^s - a_{fs}^d$ also has the added
benefit of  producing an orthogonal constraint in the $a_{fs}^s,a_{fs}^d$ plane which offers strong
prospects of $\it direct$ confirmation or refutation of the dimuon anomaly in the 2011 data set, if limitations from
the production asymmetries can be overcome. 

At this time, available combined fits indicate a 3.3$\sigma$ deviation from the SM \cite{Ligeti:2010ia,Lenz:2010gu} based on the older measurement and not incorporating the
new $\rm LHC_b$ and \DO \ measurements. We use this result in our analysis. In the model
we have studied and attempted to link to baryogenesis, one expects $h_s = h_d$ and $ \sigma_s =
\sigma_d$, and the best fit values are $h_q =0.255$ and $2\sigma_q=180^{o}+63.4^o$ based on the smaller data set.

\subsection{Collider prospects for 2HDMs with successful EWBG}

From a collider search perspective, the highly constrained mass spectrum and parameter space consistent
with EWBG is a positive feature,  since the predictions for  collider signatures are 
sharpened. The typical mass scales of the new scalar states with successful baryogenesis  fall into two
categories. When we set  $\zeta_b^0 = {\zeta'}_b^\pm \approx 0$ in our scans we find typical values $260
\lesssim \mH \lesssim 360 {\, \rm GeV}$, $350 \lesssim \mA \lesssim 550 {\, \rm GeV}$ and $300 \lesssim
m_\pm \lesssim 350 {\, \rm GeV}$ as is shown in fig \ref{hist1}.  Alternatively, if we bias the MC
search to give as large a value of $\zeta_b^0 = {\zeta'}_b^\pm$ as possible then we find typical masses $275
\lesssim \mH \lesssim 375 {\, \rm GeV}$, $475 \lesssim \mA \lesssim 575 {\, \rm GeV}$ and $400 \lesssim
m_\pm \lesssim 500 {\, \rm GeV}$. In both cases, the simultaneous best fit value for the Higgs mass is
tightly bounded around $115 \lesssim \mH \lesssim 130 {\,\rm GeV}$. Further, in both cases $\lambda_5 \sim
0$ for successful EWGB, so the new scalars, when produced, do not dominantly decay to pairs of Higgs
particles and the mass and VEV eigenstates of the new scalars  approximately coincide. This makes
discovery of these states extremely challenging when the scalar masses are above the $t \, \bar{t}$ mass threshold as we will show. SM Higgs collider
phenomenology is not significantly affected in this scenario, and we focus our discussion on the possible
collider signatures of $\mH, \mA,m_\pm$. The challenges of discovering charged scalars at LHC are well
known and are not alleviated in this model, so we focus our discussion on the neutral scalars.

\subsubsection{Tevatron Physics}

The lower end of the above mass intervals are well within the kinematic reach of the Tevatron
and the  simplest topology of interest at colliders is a resonant $s$-channel tree-level exchange of the
new scalar states.  The partonic tree-level differential cross section for $\mH,\mA$ is given by
\bea
\frac{d \, \hat{\sigma}}{d \, \hat{t}} = \frac{1}{192 \, \pi \, \hat{s}} \, \frac{1}{|\hat{\vec{p}}|^2} \, \frac{(s - m_1^2 - m_2^2) \, (s - m_3^2 - m_4^2)}{(s- \mHA^2)^2 + \GHA^2 \, \mHA^2} \, |f^{U/D}_{ij}|^2 |f^{U/D}_{kl}|^2.
\eea
We denote partonic variables with hat superscripts and the masses are labeled with $1,2$ for initial states and $3,4$ for final states.
Here the three-momenta of the initial SM states in the center of mass (CM) frame are $\hat{\vec{p}}$. The general coupling of the scalar fields to initial state flavours $i,j$
and final state flavours $k,l$ is given by
\bea
f^{U/D}_{ij} = \yUD{i} \, \left[\delta_{i,j} \,  \eta_{U/D}  
  + \eta'_{U/D} \, V_{in} \left(\yUD{n}\right)^2 \, V^\dagger_{nj} 
  +  \delta_{i,j} \, \tilde{\eta}^{'}_{U/D}  \left(\yUD{b}\right)^2 + \cdots \right] \,,
\eea
 where $\yUD{i} = \sqrt{2} m_i^{U/D}/v$.
The expression for $f_{kl}$ is identical. Considering parton distribution function (PDF) and CKM suppression, the largest effect of these exchanges at the Tevatron is expected to be on 
the production of $t \, \bar{t}$ final states. Specializing to this case, neglecting all initial state
 masses in the kinematic expressions, and taking $y_t \sim 1$, we find
the partonic cross section for $i,j$ initial states that must be convoluted with PDF's is given by
\bea
\hat{\sigma}_{ij}= \frac{1}{192\, \pi}  \, \frac{(s - 2 \, m_t^2)}{(s- \mHA^2)^2 + \GHA^2 \, \mHA^2} \, |f_{ij}|^2  \, (\zeta_t^0)^2.
\eea
It is interesting that there is an excess in the lowest mass bin between $350$ and  $400 \,{\rm
GeV}$ in the top quark invariant mass distribution.  This range approximately coincides with the masses
preferred for EWBG in our model when  ${\zeta'}_b^\pm\approx 0$. However, we find that the
maximum enhancement of the top quark invariant mass distribution in the relevant bin is too small by a
factor of $\sim 100$ for couplings consistent with particle physics constraints and successful EWBG. The
effects of tree-level exchanges of neutral scalars consistent with EWBG in other observables at the
Tevatron are similarly too small to be observed with the  current data set. 

\subsubsection{LHC Physics}

The production of the neutral scalars through the process $gg \rightarrow \SHA$ at the Tevatron and LHC
is a promising channel for this scenario. When  we bias for large values of  $\zeta_b^{'0}$ the process
$p \, p \rightarrow \SH \, b \, \bar{b}$ at LHC is also promising\cite{Trott:2010iz}. Conversely, for the parameter
sets consistent with EWBG we find {that the production cross sections for} $p \, \bar{p} \rightarrow W^{\pm} \rightarrow \Spm \, \SHA$ at the Tevatron and
$p \, p \rightarrow W^{\pm} \rightarrow \Spm \, \SHA$ at LHC are $\lesssim 10^{-3} \, {\rm fb}$.
In Table \ref{tab:collider} we display these cross sections for several 2HDMs
consistent with EWGB: five sets in which ${\zeta'}_b^\pm \approx 0$,
and five sets in which ${\zeta'}_b^\pm$, using the stronger particle physics constraints that arise
from assuming that $f=1$ in the relation (\ref{generic}) between ${\zeta'}_b^\pm$ and ${\zeta}_b^0$.
Table \ref{tab:collider2} gives the corresponding results for models that can satisfy both the
requirements of baryogenesis and the Tevatron dimuon anomaly, using the relaxed constraints from
taking $f=0.5$.

\begin{table}[h]
\begin{center}
\begin{tabular}[t]{|c|c|c|c|c|c|c|c|c|}
  \hline
  \hline
  $m_{\pm}$ & $\mH$ & $\mA$ & $\zeta_t^0$ & ${\zeta'}_b^\pm$ &
$\sigma_{(g \, g \rightarrow \SH)}$ (fb)& $\sigma_{(g \, g \rightarrow
S_{A})}$ (fb) & $\sigma_{(p \, p \rightarrow \SH \, b \, \bar{b})}$ (fb)\\
\hline
310 & 374 & 325 & -0.52 & 0 & $275 \, {}$ & $895 \, {}$ &-  \\
317 & 332 & 337 & -0.45 & 0 & $192 \, {}$ & $762 \, {}$ & -  \\
292 & 263 & 399 & 0.43 & 0 &  $239 \, {}$ & $506 \, {}$ & -  \\
307 & 300 & 407 & 0.52 & 0 & $283\, {}$ & $662 \, {}$  & - \\
315 & 283 & 462 & -0.43 & 0 & $211\, {}$ & $221 \, {}$ & -  \\
  \hline
474 & 370 & 559 & -0.52 & 0.62 &  $277 \, {}$ & $105 \, {}$ & $0.8 \, {}$   \\
412 & 304 & 497 & 0.42 & 1.16 & $279 \, {}$ & $212 \, {}$ & $7.3 \, {}$  \\
395 & 286 & 480 & -0.43 & 1.18 & $304 \, {}$ & $260 \, {}$ & $9.8 \, {}$ \\
408 & 307 & 488 & -0.44 & 1.15 & $274 \, {}$ & $236 \, {}$ & $6.7 \, {}$ \\
435 & 311 & 532 & -0.51 & 0.93 & $270\, {}$ & $141\, {}$ & $4.2 \, {}$ \\
 \hline
  \hline
\end{tabular}
\end{center}
\caption{Sample of production cross sections for models with successful EWGB, but
insufficient dimuon excess.   
We use MSTW2008 PDF's in determining the cross sections and use the renormalization scale 
$\mu = \mHA/2$ for the PDF's. The cross sections are evaluated  
for LHC at an operating energy of $7 \,{\rm TeV}$ at leading order. 
We use the results of \cite{Djouadi:2005gi,Mantry:2007ar} for the LO cross sections.
Here we have assumed the generic relation (\ref{generic}) with $f=1$ between
${\zeta'}_b^\pm$ and ${\zeta}_b^0$.}
\label{tab:collider}
\end{table}

\begin{table}[h]
\begin{center}
\begin{tabular}[t]{|c|c|c|c|c|c|c|c|c|}
\hline
\hline
$m_{\pm}$ & $m_H$ & $m_A$ & $\zeta_t^{'0}$ & $\zeta_b^{'0}$
 & $\zeta_b^{'0}/(\zeta_b^{'0})_{\DOO}$ & $\sigma_{(g \, g \rightarrow S_{H})}$ (fb) &
 $\sigma_{(g \, g \rightarrow S_{A})}$ (fb)& $\sigma_{(p \, p \rightarrow S_{H} \, b \,\bar{b})}$ (fb)\\
\hline
355 & 350 & 523 & -0.56 & 4.8 & 1.08 &  $ 312 \, {}$ & $182 \, {}$ & $16 \, {}$ \\
315 & 276 & 453 & -0.38 & 6.0 & 1.81 & $171 \, {}$ & $194 \, {}$ & $74 \, {}$ \\
388 & 352 & 554 & -0.51 & 5.9 & 1.37 & $262 \, {}$ & $107 \, {}$ & $24 \, {}$ \\
368 & 350 & 545 & -0.40 & 5.9 & 1.36 & $159 \, {}$ & $72 \, {}$ & $24 \, {}$\\
336 & 303 & 498 & -0.40 & 5.7 & 1.30 & $165\, {}$ & $124\, {}$ & $44 \, {}$ \\
\hline
\hline
\end{tabular}
\end{center}
\caption{Same as table \ref{tab:collider}, but now with sufficient dimuon excess, allowed by
mild relaxation of the relationship
	(\ref{generic}) between $\zeta_b^{'0} $ and $\zeta_b^{'\pm}$ from $f=1$ to $f=0.5$.}
      \label{tab:collider2}
\end{table}

\begin{figure}[h]
\scalebox{0.65}{\epsfig{file=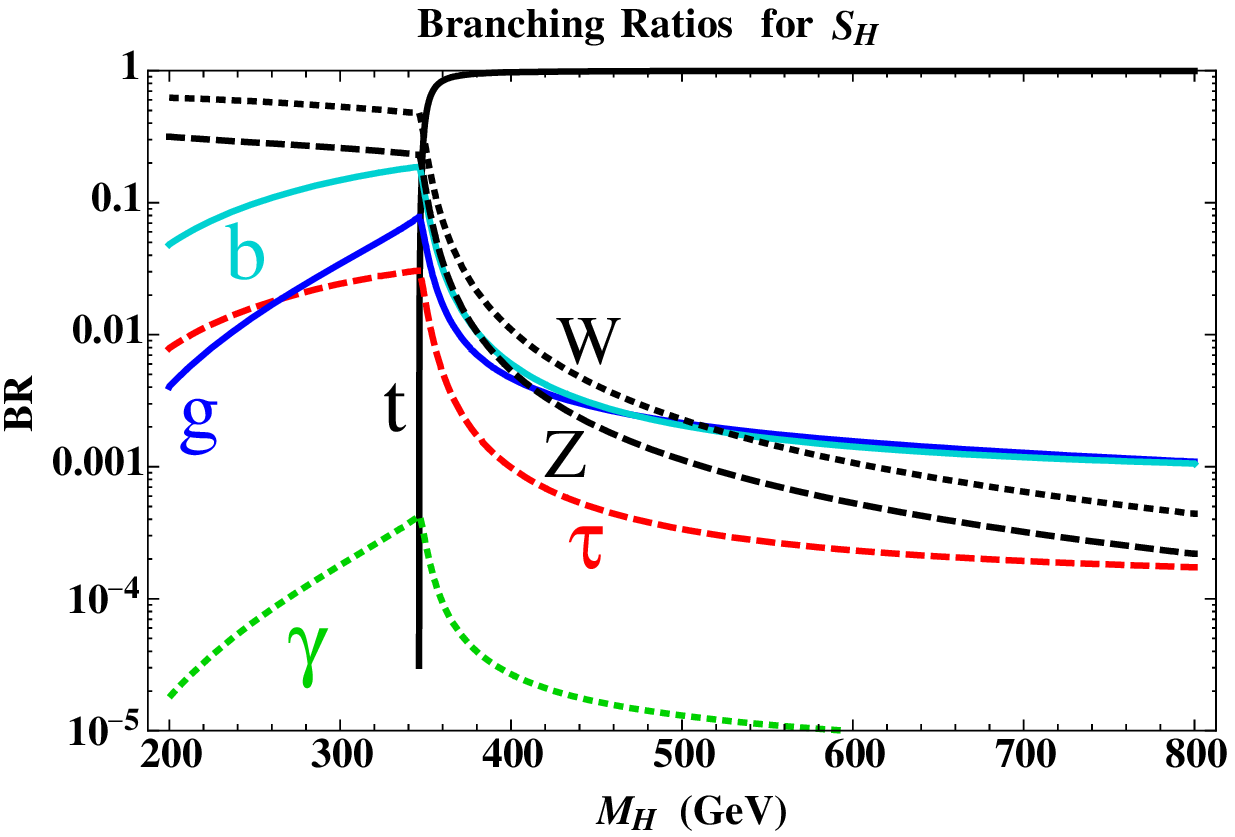}}%
\scalebox{0.65}{\epsfig{file=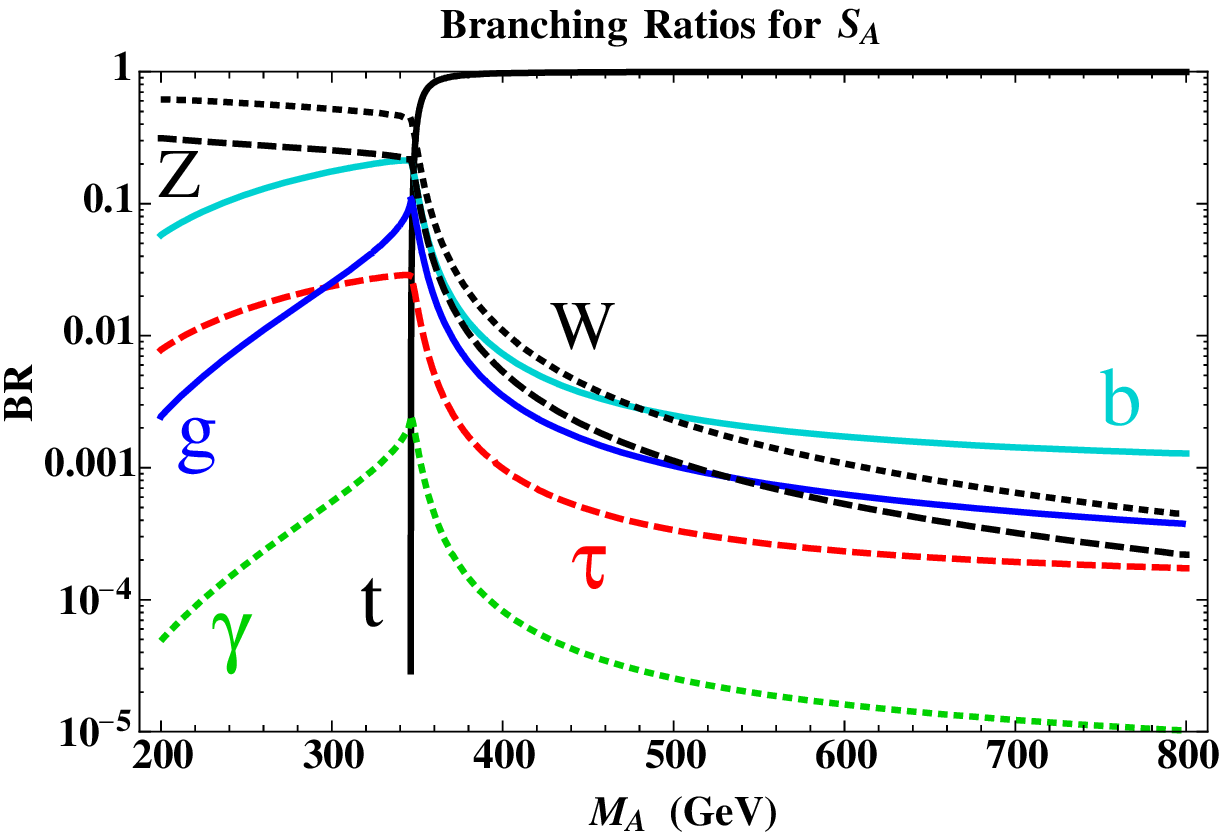}}
\caption{Branching ratios for the heavy neutral scalars. The branching is almost exclusively to $W \, W$
and $Z \, Z$ below the top quark threshold and to  $t \, \bar{t}$ above the top threshold. Here we have
neglected flavour-changing decays and decays to $hh$, the later being highly suppressed compared to
the leading decay to $W \, W$ and $Z \, Z$ or $t \, \bar{t}$ and challenging to discover. 
Each curve is labeled by the final state particle $X$ for the decay $\SHA \to X\bar X$.
We have taken $\zeta_t^0 = 0.5$, $m_t = 173.1$ and
$\zeta_b^0 = 2 m_b^2/v^2$. Flavour changing decays to $t \, \bar{c}$ or $t \, \bar{u}$ require an
insertion of $2 m_b^2/v^2$ and further CKM suppression. They are given essentially by CKM suppressed
rescalings of the shown ${\rm Br}(b \, \bar{b})$.}
\label{branching}
\end{figure}

For the parameter sets consistent with EWBG, we find that the production through gluon fusion indicates
that LHC's current run (with $\gtrsim 1 {\rm fb}^{-1}$ on tape) should have produced hundreds of events.
The branching ratios of the neutral scalars are determined using the results summarized in \cite{Djouadi:2005gi}.
Recall our convention that the field $S$ does not get a VEV,
because of this and the fact that  $|\lambda_5| \ll 1$, the mass eigenstate fields only get small VEVs of the order
\bea
\langle \SH' \rangle = v \epH = \frac{v^3 \LfiveR}{\mH^2 - m_h^2}, \quad  \quad \langle \SA' \rangle = v \epA = \frac{v^3 \LfiveI}{\mA^2 - m_h^2}. \nn 
\eea
When EWBG occurs, these vacuum expectation values are $\langle \SH' \rangle \simeq \langle \SA' \rangle \simeq 10 \, {\rm GeV}$.
This significantly suppresses decays to $WW$, $ZZ$ making these scalars difficult to discover at LHC. 
The largest decay widths are typically to fermions with
\bea
\Gamma(\SHA \rightarrow f \, \bar{f}) = \frac{G_F \, N_C}{8 \, \sqrt{2} \, \pi}  \, v^2 \, |\zeta_f^0|^2 \, \mHA  \, \beta_f^n,
\eea
and $\beta = (1 - 4 m_f^2/\mHA^2)^{1/2}$. Here $n = 3$ for $\mH$ and $n=1$ for $\mA$. The neutral scalars
decay almost exclusively into $t \, \bar{t}$ final states when they are above the $t \, \bar{t}$
threshold. Other phenomenologically interesting final states due to their clean collider signatures are
$\gamma \, \gamma$, $ZZ$, $WW$ and $\tau^+ \, \tau^-$. We show in Fig.\ (\ref{branching}) the dominant branching ratios as
a function of the scalar masses for typical parameter values. We also show the branching ratios to EW
final states that are potentially promising for discovery.

Below the $t \, \bar{t}$ threshold, strong discovery prospects exist through decays to $W \, W$ and $Z \,
Z$ using standard searches. Typically in these scenarios at least one heavy neutral scalar is below the
$t \, \bar{t}$ threshold. However above the $t \, \bar{t}$ threshold, discovery prospects are much
dimmer. When $\mHA > 2 \, m_t$ large numbers of resonant $t \, \bar{t}$ pairs are produced. Unfortunately
the $t \, \bar{t}$ production from QCD is far larger because the signal process is loop-suppressed.  For
example, for $\SA\to t \, \bar{t}$ with a mass scale of $\mA \sim 500  \, {\rm GeV}$, the 
approximate cross section for $\sigma(g \, g \rightarrow \SA)$ is $0.2 \,{\rm fb}$ while the QCD $t \,
\bar{t}$ background with $M_{t \, \bar{t}} \sim 500  \, {\rm GeV}$  is given by $\sim 10 \, {\rm pb}$ at
leading order. 
 The first measurements of the $t \, \bar{t}$ invariant mass spectrum at LHC have recently appeared
\cite{atlasnote}. However it is unlikely that the resonant production of  $t \, \bar{t}$ due to these
heavy scalars can be uncovered in the $t \, \bar{t}$ invariant mass spectrum  for quite some time.
Precision studies would have to be performed, possibly using the angular distributions of the
reconstructed  $t \, \bar{t}$ pair since this signal is through a scalar resonance.  After event selection
cuts, the reconstructed partonic $t \, \bar{t}$ spectra has only 5 events in the  $M_{t \, \bar{t}} \sim
500  \, {\rm GeV}$ mass region in this study. Thus total event selection cut suppression of the
background and signal is $\sim 10^{-3}$.

Due to the difficulty of overcoming the $t \, \bar{t}$ QCD background, and the lack of efficient light flavour tagging at LHC to uncover, for example, CKM suppressed $t \, \bar{c}$ decays, utilizing a highly suppressed branching 
ratio such as to $\tau^+ \, \tau^-$ might be considered more hopeful. The dominant  
background\footnote{This is most difficult to significantly reduce with selection cuts. A background of 
$t \, \bar{t}$ also exists with $\tau$ pairs produced in the decay
of the tops. The latter should be more reducible due to its associated missing energy and multi-jet 
topology.} in this case is the inclusive process $p \, p \rightarrow Z \, X \rightarrow \tau^+ \, 
\tau^- \, X $, whose tree-level electroweak contribution has an antiquark initial-state PDF suppression at LHC.
The partonic cross section for this background is given by
\bea
\hat{\sigma}_{f} = \frac{\hat{s} \, (4 \pi)^2 \, \alpha^2_{EW}}{48 \, \pi \, \sin^4\theta_W\, \cos^4\theta_W} \, \frac{\left[(g_V^f)^2 + (g_A^f)^2 \right] \, \left[(g_V^\tau)^2 + (g_A^\tau)^2 \right]}{(\hat{s} - m_z^2)^2 + m_z^2 \, \Gamma_Z^2}
\eea
Here $g_V,g_A$ are the usual electroweak couplings that depend on the isospin of the initial state fermions $f$. We use $\Gamma_Z = 2.5 \, {\rm GeV}$ and $m_z = 91.2\, {\rm GeV}$. 
We use hat superscripts to denote partonic quantities.
The expression we use to implement cuts on the $2 \rightarrow 2$ processes of interest,
in ${\rm fb}$ units, is given by
\bea
\frac{d \sigma}{d M_{\tau \tau}} = \frac{0.3894 \times 10^{12} \, M_{\tau \, \tau}}{s_{had}} \int_{-y_B^0}^{y_B^0} d \, y_B  \int^{z_0}_{-z_0} d \, z \, \sum_{f} f_f(R \,  e^{y_B}) \, f_{\bar{f}}(R \,  e^{-y_B}) \, \hat{\sigma}_{f} 
\eea
where $R = \sqrt{M_{\tau \, \tau}^2/s_{had}}$, $f_{f},f_{\bar{f}}$ are the 2008 MSTW parton densities of the initial state quarks, $s_{had}$ is the hadronic center of mass energy, and $Y_B$ is the boost rapidity of the subprocess frame. We treat the highly boosted $\tau$'s as $\tau$ jets
required to have rapidity $|y_{1,2}| < Y_{max}$ with cuts on $p_T$ of $p_T^{min}$ so that
\bea
y_B^0 &=& {\rm Min}[Y_{max} -{\rm Log}[M_{\tau \, \tau}^2/s]/2], \nn \\
z^0 &=&  {\rm Min}[\sqrt{1 - 4 \, (p^{min}_T)^2/M_{\tau \, \tau}^2}, \tanh(Y_{max} - |y_B|)].
\eea
This result for hadronic cuts on $2 \rightarrow 2$ processes is consistent with \cite{Lane:1991qh}.
We find that the $d \, \sigma/ d M_{\tau \tau}$ invariant mass spectrum has a peak of resonantly produced $\tau^+ \, \tau^-$ final states typically 
$10^{-3}$ the background spectrum when a $p_T$ cut of $10 \, {\rm GeV}$ is applied to the leptons and the $\tau$ jets are restricted to have a rapidity of $< 3$.
Thus discovery in this channel is unlikely. Above the $t \, \bar{t}$ threshold, relying on decays to $W \, W$ and $Z \, Z$ offers some prospects for discovery.

\section{Discussion and conclusions}
\label{conclusions}

In this paper we have given an exhaustive account of EWBG in two-Higgs
doublet models.  Our study, while focusing on MFV Yukawa couplings and a
simultaneous explanation of the \DO \,  dimuon excess, is general and
encompasses earlier popular choices of Yukawa couplings, such as type I and II
models.  The MFV two Higgs doublet model also offers a controlled theoretical 
framework where new CP violating phases can appear at the EW scale without
introducing inconsistencies with a multitude of other flavour measurements. 
Unlike previous works, we have considered the most general set of
couplings in the Higgs potential and have made no simplifying assumptions to
reduce the parameter space.  Instead we used a Markov Chain Monte Carlo to explore
the full space of models, aiming toward those with a strong first order phase
transition and large CP violation in the bubble walls.  

Despite this, we find it more difficult to
obtain a large enough baryon asymmetry than previous studies indicated.  This is
partly because we have imposed more phenomenological constraints; in addition to
accelerator mass bounds and the neutron EDM that were considered in previous studies,
we included  EWPD, $R_b$, $b\to s\gamma$ and perturbativity of couplings
up to a minimum Landau pole scale of $1-2$ TeV.  
In addition, we have avoided simplifying assumptions about the shape of the bubble wall profile that
artificially increased the baryon asymmetry in previous work. For every model in
the chain from the Monte Carlo, we numerically determined the actual solutions for
the bubble wall profiles and the CP-violating phase $\theta(z)$ that sources baryogenesis. 
 In the process we elucidated a dynamical mechanism by which $\theta(z)$ tends to be
suppressed: the existence of nonzero $\theta(z)$ depends upon exciting the field
$S$ whose VEV is zero at $T=0$.  Thus there is finite-temperature suppression factor inherent
in $\theta(z)$ which would be overlooked in a seemingly reasonable guess for its
profile.   We numerically solved the Boltzmann transport equations for each
profile to determine the baryon asymmetry.

The effective potential approach we have employed has theoretical uncertainties
that are difficult to quantify without going through the arduous step of adding
two-loop corrections.  Ultimately a lattice study would be welcome for verifying
the details of the phase transition properties, but this is an even more ambitious
undertaking.\footnote{Ref.\ \cite{Patel:2011th} has also promoted a
method to extract $v_c/T_c$ from the effective potential in a
gauge-invariant manner.}\  
A more tractable improvement on our treatment would be to rederive
the transport equations and related coefficients, which we have taken from ref.\
\cite{FHS} without modification.  In particular, it is possible that including
extra species in the network of equations could enhance the baryon asymmetry.
Tau leptons, although weakly coupled, diffuse farther than other species, and
could thus carry the CP asymmetry deeper into the symmetric phase and boost the
integral (\ref{etabint}) determining $n_B$.\footnote{We thank S.\ Tulin for pointing
out this possibility}\  A crucial issue to resolve concerns the correct form of
the source term in the transport equations.  Ref.\ \cite{Tulin:2011wi} has recently claimed
that 2HDMs can more readily produce a large enough baryon asymmetry (as well as
address $B$ meson anomalies including the one we have considered) than we find.
This is in part due to the use of a source term that is first order in
derivatives, rather than the semiclassical force term with two derivatives that we
have adopted.

\begin{figure}[t]
\scalebox{0.65}{\epsfig{file=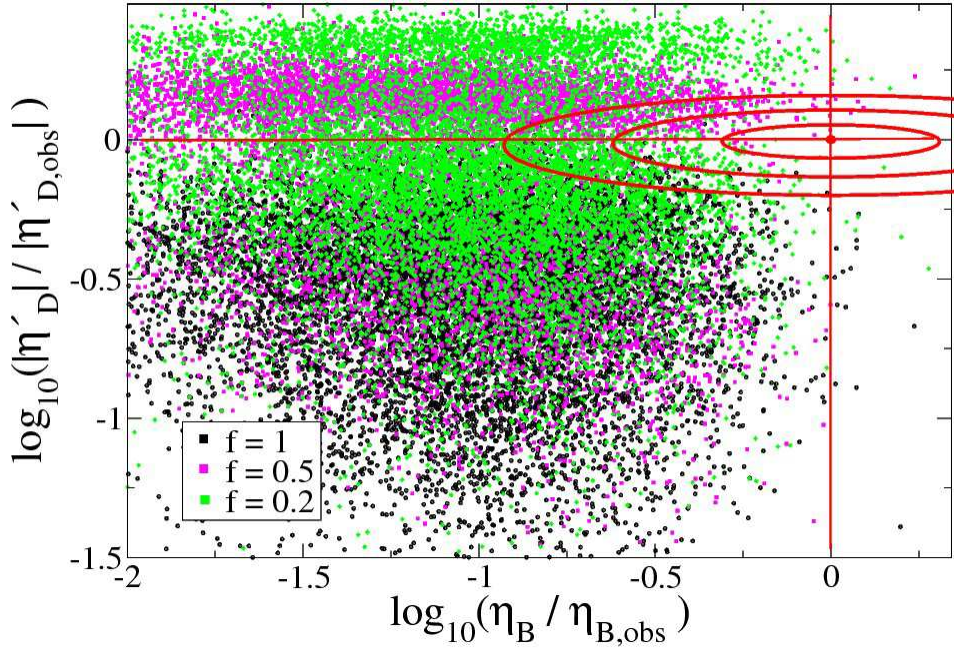}}%
\caption{Scatter plot of $|\eta'_D|$ normalized to the observed value
(as needed to explain the \DO\ dimuon excess) versus the baryon asymmetry
relative to its observed value.  The 1, 2 and 3-$\sigma$ error ellipses as
described in the text are shown.  Black points are for $f=1$, magenta for $f=0.5$,
green for $f=0.2$, referring to $f$ defined in eq.\ (\ref{generic}).}
\label{scatter}
\end{figure}

Given these theoretical uncertainties in computing the baryon asymmetry, it
seems reasonable to believe that the true prediction for a given model could
differ by a factor of 2 from our determinations, despite the pains we have taken
to get accurate results.  At the same time, there is uncertainty in the value
of the new Yukawa coupling parameter $\eta'_D$ needed to match the \DO\ anomaly; from
table 17 of \cite{Lenz:2010gu}, we can estimate that the 1-$\sigma$ error band in 
$h_q$ of eq.\ (\ref{hqeq}) is $h_q = 0.258 {+0.63\atop -0.80}$ using the method
advocated in \cite{barlow}.  Since $h_q\sim (\eta'_D)^2$, this translates to the 
relative error $\delta\eta'_D/|\eta'_D| ={+0.122\atop -0.154}$.
From these considerations we can give a better picture of the tension
between explaining the dimuon excess and baryogenesis as shown in fig.\
\ref{scatter}.  It will be important to reduce the systematic errors in the
prediction of the baryon asymmetry to know how serious the apparent tension really is.

One advantage of having a small allowed region of parameter space is that
we can make somewhat distinctive predictions for collider searches.
Typically the discovered mass spectrum is a light Higgs, and one or more heavy
scalars  below the $t \, \bar{t}$ mass threshold with branching ratios as
shown in fig.\ \ref{branching}. LHC can discover such a spectrum in the near future. This scenario for EWBG is  falsifiable for
most of the  remaining allowed parameter space.

\acknowledgments
JC thanks the particle theory group of the Perimeter Institute for its hospitality
while this work was started, H.\ Logan and D.\ Neilsen for helpful 
correspondence or discussions, and the organizers and participants of
``Electroweak Baryogenesis in the Era of the LHC'' at the Weizmann
Institute for stimulating discussions.
MT thanks Mark Wise for discussions and collaboration related to material
presented here and Pierre Savard for helpful communication.
We thank Pietro Slavich for pointing out an error in the formula for 
$R_b$ in the original manuscript.

\appendix

\section{Counterterms}
\label{ctapp}

Denoting derivatives of the Coleman-Weinberg potential $V_{\rm CW}$,
evaluated at the tree-level VEVs, by $V_{1,x}$ and $V_{1,xy}$,
\bea
	\delta\lambda\, v^2 &=& \sfrac12(\sfrac{1}{v}V_{1,h} - V_{1,hh})\\
	\delta \mu^2 &=& \sfrac12(\sfrac{3}{v}V_{1,h} - V_{1,hh})\\
	\delta\LthreeR\, v^2 &=& 
	\sfrac12(-V_{1,\SR\SR} + V_{1,\SI\SI})\\
	\delta\LthreeI\, v^2 &=& V_{1,\SR\SI}\\
	\delta m_1^2 &=& -\sfrac12(V_{1,\SR\SR} + V_{1,\SI\SI})\\
	\delta m_{2R}^2 &=& \sfrac12( V_{1,h\SR} - \sfrac{3}{v} V_{1,\SR})\\
	\delta m_{2I}^2 &=& \sfrac12(-V_{1,h\SI} + \sfrac{3}{v} V_{1,\SI})\\
	\delta\LfiveR\, v^2 &=& -V_{1,h\SR} + \sfrac{1}{v}V_{1,\SR}\\
	\delta\LfiveI\, v^2 &=& -V_{1,h\SI} + \sfrac{1}{v}V_{1,\SR}
\eea
As described in the text, the Goldstone boson contributions to 
$V_{CW}$ must be IR-regulated in order to give meaningful results.

\section{Field-dependent masses}
\label{masses}

The neutral components of the complex scalar doublets can be written
as
\beq
	H = {1\over\sqrt{2}}(h + i h_I),\quad
	S = {1\over\sqrt{2}}(\SR + i \SI)
\eeq
In our evaluation of the effective potential we are free to make a global SU(2)$_L$ transformation that sets $h_I=0$ so that $V_{\rm eff}$ depends only upon $h,\SR,\SI$.  The resulting $4\times 4$ neutral mass matrix, including finite-$T$ Debye mass corrections, has components given by
\bea
V_{h,h} &=& 
\frac{\lambda}{4}\left(3{h}^{2}-{v}^{2}\right)
+  \sfrac12(\lambda_1+\lambda_2)(\SRsq + \SIsq)
+{\lambda_3}\left(\SRsq-\SIsq\right) 
+3h({\LfiveR}\SR + {\LfiveR}\SI)\nonumber\\
&+& (3\lambda+4\lambda_1+2\lambda_2+6\yht^2 +\sfrac92 g^2 + \sfrac32
g'^2)\sfrac{T^2}{24}\\
V_{h_I,h_I} &=& 
\frac{\lambda}{4}\left({h}^{2}-{v}^{2} \right)
+\sfrac12(\lambda_1+\lambda_2)(\SRsq + \SIsq)
-{\lambda_3}\left(\SRsq-\SIsq\right) 
+\phantom{3}h({\LfiveR}\SR + {\LfiveR}\SI)\nonumber\\
&+&(3\lambda+4\lambda_1+2\lambda_2+6\yht^2+\sfrac92 g^2 + \sfrac32
g'^2)\sfrac{T^2}{24}\\
V_{\SR,\SR} &=& {m_1^2}
+\sfrac12\,({\lambda_1}+{\lambda_2} + 2\lambda_3)\,{h}^{2}
+3{\LfourR}\,h\SR -{\LfourI}\,h \SI
+{\lambda_6}\,(3\SRsq+\SIsq)\nonumber\\
&+&(4\lambda_1+2\lambda_2+12\lambda_6+6|\eta_t|^2+\sfrac92 g^2 + \sfrac32
g'^2)\sfrac{T^2}{24}\\
V_{\SI,\SI} &=& {m_1^2}+\sfrac12\,({\lambda_1}+{\lambda_2}-2{\lambda_3}
)\,{h}^{2}-3{\LfourI}\,h\SI+{\LfourR}\,h\SR+{\lambda_6}\,
(3\SIsq+\SRsq)\nonumber\\
&+&(4\lambda_1+2\lambda_2+12\lambda_6+6|\eta_t|^2+\sfrac92 g^2 + \sfrac32
g'^2)\sfrac{T^2}{24}\\
V_{h,h_I} &=& \,{\lambda_3}\,\SR\SI+h({\LfiveR}\,\SI-{
\LfiveI}\,\SR)\\
V_{h,\SR} &=&{m_{2R}^2}+
({\lambda_1}+{\lambda_2} + 2\lambda_3)\,h\SR + \sfrac12{\LfourR}
\left( 3\SRsq+\SIsq \right) 
-{\LfourI}\SR\,\SI
+\sfrac32\,{\LfiveR}{h}^{2}\nonumber\\
&+& 6( \LfourR+\LfiveR + \yht|\eta_t|\cos{\varphi_\eta})\sfrac{T^2}{24}\\
V_{h,\SI}  &=&-{m_{2I}^2}
+(\lambda_1+\lambda_2-2\lambda_3) h \SI
-\sfrac12 {\LfourI}\left(\SRsq+3\SIsq \right) 
+{\LfourR} \SR\,\SI
+\sfrac32\,{\LfiveI}\,{h}^{2}\nonumber\\
&+& 6( \LfiveI-\LfourI - \yht|\eta_t|\sin{\varphi_\eta})\sfrac{T^2}{24}\\
V_{h_I,\SR}  &=&
{m_{2I}^2}+2\,{\lambda_3}\,h\SI+ 
\sfrac12{\LfourI}\left(3\SRsq+\SIsq \right) 
+{\LfourR}\SR\, \SI
-\sfrac12\,{\LfiveI}\,{h}^{2}\\
V_{h_I,\SI} &=&
{m_{2R}^2}+ 2\,{\lambda_3}\,h\SR
+ \sfrac12{\LfourR}\left(\SRsq+3\SIsq \right) 
+{\LfourI}\,\SR\,\SI
+\sfrac12\,{\LfiveR}\,{h}^{2}\nonumber\\
&+& 6( \LfourR+\LfiveR + \yht|\eta_t|\cos{\varphi_\eta})\sfrac{T^2}{24}\\
V_{\SR,\SI}  &=& h({\LfourR}\,\SI\,-{\LfourI}\,\SR)+2\,{\lambda_6}\,\SI\,\SR
\eea
Here $\eta_t \equiv |\eta_t| e^{i\varphi_\eta}$ and we have denoted the real and imaginary parts of complex masses and couplings by $m_2^2 = m^2_{2R} + i m^2_{2I}$, {\it etc.}  The mass matrix must be numerically diagonalized to find the mass squared eigenvalues.

\noindent
The charged scalar mass matrix elements are given by
\bea
	V_{h^+,h^-} &=& {\lambda\over 4}(h^2 - v^2)+\sfrac12\lambda_1
	(\SRsq + \SIsq) + h(\LfiveR \SR + \LfiveI \SI)\nonumber\\
	&+& (3\lambda+4\lambda_1+2\lambda_2+6 \yht^2+\sfrac92 g^2 
	  + \sfrac32 g'^2)\sfrac{T^2}{24}\\
	V_{h^+,s^-} &=& (m_2^2)^* +\sfrac12 \lambda_2 h(\SR + i \SI)
	+ \lambda_3 h(\SR-i\SI) +\sfrac12\lambda_4^*(\SRsq+\SIsq)
	+\sfrac12\lambda_5 h^2\nonumber\\
	&+&6 (\lambda_5 + \lambda_4^* + \yht\eta_t^*)\sfrac{T^2}{24},
\eea	
\bea
	V_{s^+,h^-} &=& V_{h^+,s^-}^* \\
	V_{s^+,s^-} &=& m_1^2 + \sfrac12\lambda_1 h^2 + 
	h(\LfourR \SR - \LfourI \SI) 
	+ \lambda_6(\SRsq - \SIsq)\nonumber\\
	&+&(4\lambda_1+2\lambda_2+12\lambda_6+6|\eta_t|^2+\sfrac92 g^2 
	 + \sfrac32 g'^2)\sfrac{T^2}{24} \,,
\eea
and the charged scalar mass eigenvalues are 
\beq
m^2_\pm = \frac12(V_{h^+,h^-} + V_{s^+,s^-})
\pm \frac12\sqrt{(V_{h^+,h^-} - V_{s^+,s^-})^2 + 4|V_{h^+,s^-}|^2} \,.
\eeq
\\
\noindent The gauge boson masses in the basis of $W_i^\alpha,B^\alpha$ are given by
\bea
	m^2_{W_i^\alpha W_j^\beta} &=& 
	\left(\sfrac14 g^2\phi^2 + 2 g^2 T^2 \delta_{\alpha,\|}
	\right)\delta_{ij}\delta_{\alpha\beta}\\
	m^2_{W_i^\alpha B^\beta} &=& 
	\left(\sfrac14 gg'\phi^2 \right)\delta_{\alpha\beta}\\
	m^2_{B^\alpha B^\beta} &=& 
	\left(\sfrac14 g'^2\phi^2
	+ 2 g'^2 T^2 \delta_{\alpha,\|} \right)\delta_{\alpha\beta}
\eea
Here $\phi^2 = h^2 + \SRsq + \SIsq$ denotes the total electroweak-breaking VEV and $\delta_{\alpha,\|}$ indicates that only the 
longitudinal polarizations get a Debye correction at this order. \\

\noindent Finally, the top quark mass is given by
\beq
	m_t^2 = \sfrac12 \left(\left(\yht h 
+ |\eta_t|(\SR\cos\varphi_{\eta}-\SI\sin\varphi_{\eta})\right)^2 
   + |\eta_t|^2(\SR\sin\varphi_\eta+\SI\cos\varphi_\eta)^2\right) \,.
\eeq

\section{Beta functions}
\label{app:betafns}
The scaled beta functions, generalizing the results of ref.\ \cite{Ferreira:2009jb},
are as follows.  We have defined $\lambda^{R,I}_n$ to be the real or
imaginary part, respectively, of $\lambda_n$.  By ``scaled,'' we mean
that $\beta_i = \hat\beta_i/16\pi^2$ in terms of the 
conventionally normalized function, $\beta_\lambda =
d\lambda/d\ln\mu$, where $\mu$ is the renormalization scale.
\bea
\hat\beta_{\lambda} &=& 6 \, \lambda^2 + 8 \lambda_1^2 + 8 \lambda_1 \, \lambda_2 + 4 \lambda_2^2 + 16 \lambda_3^2 + 48 |\lambda_5|^2 + \frac{3}{2} \left(3 g^4 + g'^4 + 2 g^2 \, g'^2 \right) 
\nonumber\\ &\,& - 3 \lambda \left(3 g^2 + g'^2 - 4 \yht^2 \right)
 - 12 \, \yht^4, \nn \\
\hat\beta_{\lambda_1} &=& \left(\lambda + 4 \, \lambda_6 \right)
(3 \lambda_1 + \lambda_2) + 4 \lambda_1^2 + 2 \lambda_2^2 + 
8 \lambda_3^2 + 4 |\lambda_5|^2 + 16 \left(\LfiveR \, 
\LfourR + \LfiveI \, \LfourI  \right) + 4 |\lambda_4|^2,
  \nn \\ &\,& + \frac{3}{4} \left(3 g^4 + g'^4 - 2 g^2 g'^2 \right) 
- 3 \lambda_1 \left(3 g^2 + g'^2 - 2 \yht^2 - 2 |\eta_t|^2 - 
2 |\eta_b|^2 \right) - 12 \yht^2 \, |\eta_t|^2, \nn \\ 
\hat\beta_{\lambda_2} &=& \left(\lambda + 4 \, \lambda_6 \right) 
\lambda_2 + 8 \, \lambda_1 \, \lambda_2 + 4 \lambda_2^2 + 32 \, 
\lambda_3^2 + 10 |\lambda_5|^2 + 4 \left(\LfiveR \,
 \LfourR + \LfiveI \, \LfourI  \right) 
+ 10 |\lambda_4|^2, \nn \\ 
&\,& + 3 \, g^2 g'^2 - 3 \lambda_2 \left(3 g^2 + g'^2 - 2 \yht^2 - 2 |
\eta_t|^2  - 2 |\eta_b|^2 \right) - 12 \yht^2 |\eta_t|^2,\nn 
\eea
\bea
\hat\beta_{\lambda_3} &=&  \left(\lambda + 4 \, \lambda_6 \right) \lambda_3 + 8 \, \lambda_1 \, \lambda_3 + 12 \lambda_2 \, \lambda_3 + 5 \, |\lambda_5|^2 + 2 \left(\LfiveR \, \LfourR + \LfiveI \, \LfourI  \right) + 5 |\lambda_4|^2, \nn \\
&\,& - 3 \lambda_3 \left(3 g^2 + g'^2 - 2 \yht^2 - 2 |\eta_t|^2 - 2 |
\eta_b|^2 \right) - 6 \yht^2 |\eta_t|^2, \nn \\
\hat\beta_{\lambda_4}^R &=& 24 \lambda_6 \, \LfourR + 6 \lambda_1 (\LfiveR + \LfourR) + 4 \lambda_2 \, \LfiveR + 8 \lambda_2 \, \LfourR + 4 \lambda_3 \, \LfiveR \ + 20 \lambda_3 \, \LfourR 
\nn\\ &\,& - 3 \, \LfourR \left(3 g^2 + g'^2 - \yht^2 - 
3 |\eta_t|^2 - 3 |\eta_b|^2 \right) 
 - 12 \left((\eta_t^R)^3 + (\eta_t^I)^2 \, (\eta_t^R) \right) \, \yht,
\nn \\ 
\hat\beta_{\lambda_4}^I &=& 24 \lambda_6 \, \LfourI + 6 \lambda_1 ( - \LfiveI + \LfourI) - 4 \lambda_2 \, \LfiveI + 8 \lambda_2 \, \LfourI - 4 \lambda_3 \, \LfiveI \, - 20 \lambda_3 \, \LfourI 
 \nn\\ &\,& - 3 \, \LfourI \left(3 g^2 + g'^2 - \yht^2 - 3 |
\eta_t|^2 - 3 |\eta_b|^2 \right) 
 - 12 \left((\eta_t^I)^3 + (\eta_t^R)^2 \, (\eta_t^I) \right) \, \yht, 
\nn \\
\hat\beta_{\lambda_5}^R &=& 6 \lambda \, \LfiveR + 6 \lambda_1 (\LfiveR + \LfourR) + 8 \lambda_2 \, \LfiveR + 4 \lambda_2 \, \LfourR + 20 \lambda_3 \, \LfiveR \ + 4 \lambda_3 \, \LfourR 
 \nn\\&\,& - 3 \, \LfiveR \left(3 g^2 + g'^2 -  3 \yht^2 -  |\eta_t|^2 -  
|\eta_b|^2 \right)
 - 12  \, \eta_t^R \, \yht^3,
\nn \\ 
\hat\beta_{\lambda_5}^I &=& 6 \lambda \, \LfiveI + 6 \lambda_1 (\LfiveI - \LfourI) + 8 \lambda_2 \, \LfiveI - 4 \lambda_2 \, \LfourI - 20 \lambda_3 \, \LfiveI  + 4 \lambda_3 \, \LfourI 
\nn\\ &\,& - 3 \, \LfiveI \left(3 g^2 + g'^2 -  3 \yht^2 -  |\eta_t|^2 -  
|\eta_b|^2 \right)
 - 12  \, \eta_t^I \, \yht^3,
\nn \\ 
\hat\beta_{\lambda_6} &=& 24 \lambda_6^2 + 2 \lambda_1^2 + 2 \lambda_1 \,
\lambda_2 + \lambda_2^2 + 4 \lambda_3^2 + 12 \lambda_1^2 + \frac38 \left(3 g^4 + g'^4 + 2 g^2 g'^2 \right)
\nn\\ &\,& - 3 \lambda_6 \left(3 g^2 + g'^2 - 4 |\eta_t|^2 -
 4 |\eta_b|^2\right)
- 6 |\eta_t|^4 - 6 |\eta_b|^4.
\eea

\end{document}